\newcommand*{\ATLASLATEXPATH}{latex/}

\pdfoutput=1

\PassOptionsToPackage{english}{babel}
\documentclass[cernpreprint, texlive=2016, UKenglish]{\ATLASLATEXPATH atlasdoc}
\DeclareOldFontCommand{\rm}{\normalfont\rmfamily}{\mathrm}
\DeclareOldFontCommand{\sf}{\normalfont\sffamily}{\mathsf}
\DeclareOldFontCommand{\tt}{\normalfont\ttfamily}{\mathtt}
\DeclareOldFontCommand{\bf}{\normalfont\bfseries}{\mathbf}
\DeclareOldFontCommand{\it}{\normalfont\itshape}{\mathit}
\DeclareOldFontCommand{\sl}{\normalfont\slshape}{\@nomath\sl}
\DeclareOldFontCommand{\sc}{\normalfont\scshape}{\@nomath\sc}

\usepackage{array,booktabs}
\usepackage{\ATLASLATEXPATH atlaspackage}
\usepackage{\ATLASLATEXPATH atlasbiblatex}
\usepackage{\ATLASLATEXPATH atlascontribute}
\usepackage{\ATLASLATEXPATH atlasphysics}
\usepackage{rotating}

\graphicspath{{logos/}{figures/}}

\usepackage{HIGG-2016-27-PAPER-defs}
\usepackage{multirow}
\usepackage{verbatim}
\usepackage[utf8]{inputenc} 
\usepackage{lineno, blindtext, booktabs}
\usepackage{array, siunitx}
\usepackage{placeins}
\AtlasTitle{Search for Higgs boson pair production in the $b\bar{b}WW^{\ast}$
decay mode at $\sqrt{s}$=13 TeV with the ATLAS detector}

\AtlasAbstract{
  A search for Higgs boson pair production in the
$b\bar{b}WW^{\ast}$ decay mode is performed in the $b\bar{b}\ell\nu qq$
final state using 36.1 fb$^{-1}$ of proton-proton collision data at a centre-of-mass energy of 13 TeV
recorded with the ATLAS detector at the Large Hadron Collider.
No evidence of events beyond the background expectation is found. 
Upper limits on the  non-resonant $pp
\to HH$  production cross section of 10 pb   and  
on the resonant production  cross section  as a function of the $HH$ invariant mass are 
obtained. Resonant production limits are set for scalar and 
spin-2 graviton hypotheses in the mass range 500 to 3000 GeV. 
}

\author{The ATLAS Collaboration}
\AtlasRefCode{HIGG-2016-27}
\PreprintIdNumber{CERN-EP-2018-237}
\AtlasDate{\today}
\AtlasJournal{JHEP}
\hypersetup{pdftitle={ATLAS document},pdfauthor={The ATLAS Collaboration}}

\begin{document}

\maketitle
\tableofcontents
\clearpage
\section{Introduction} \label{introduction}

The Higgs boson ($H$) is an essential part of the Standard Model (SM) and it
has a crucial role  in the electroweak symmetry breaking (EWSB)
mechanism~\cite{PhysRevLett.13.321,HIGGS1964132,PhysRevLett.13.508,PhysRevLett.13.585,PhysRev.145.1156,PhysRev.155.1554}. In this mechanism, an SU(2) doublet bosonic scalar field  is
subject to a potential energy term whose shape allows the doublet
field to acquire a vacuum expectation value that breaks the SU(2) symmetry and
produces the Higgs boson and its
potential energy term. This potential is the last piece of the SM Lagrangian which is yet to be
directly tested. 

\begin{figure}[!h]
\begin{center}
\includegraphics*[width=0.95\textwidth] {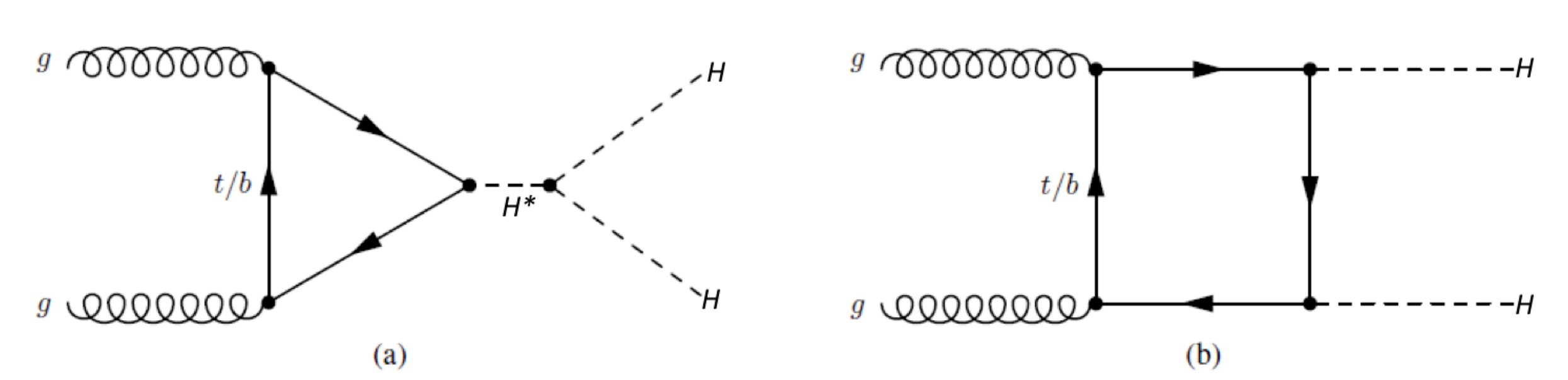}
\caption{Leading-order Feynman diagrams for non-resonant production 
  of Higgs boson pairs in the Standard Model through (a) the Higgs 
  boson self-coupling and (b) the Higgs--fermion Yukawa
  interaction. The $H^*$ refers to the off-shell Higgs boson
  mediator. }
  \label{fig:nr_hh_production}
\end{center}
\end{figure}

The shape of the Higgs boson potential in the SM can be expressed as a
function of the Fermi coupling constant $G_F$ and the Higgs boson mass $m_H$.
A direct phenomenological prediction of the SM
due to the potential is the interaction of the Higgs boson with itself at 
tree level (self-interaction), which can be probed by studying
di-Higgs boson production in proton--proton collisions, as illustrated in
Figure~\ref{fig:nr_hh_production}(a). The self-interaction 
diagram together with the quark-loop contributions, primarily via the top-Higgs Yukawa coupling, Figure~\ref{fig:nr_hh_production}(b), 
are the leading-order Feynman diagrams for Higgs boson pair production. 
The SM cross section for $pp \to HH$ is extremely small, e.g. 33.4~fb at 13~\TeV\ \cite{deFlorian:2016spz}. 

Physics beyond the SM can manifest in the increased production with
respect to the SM predictions of the non-resonant $HH$ final state
or in the resonant production of particles that decay into a pair of SM Higgs bosons.
The analysis presented here is potentially sensitive to cases where the decaying particle is a scalar, as in the MSSM~\cite{Haber:1984rc} and 
2HDM models \cite{BRANCO20121}, or a spin-2 graviton, as in 
Randall--Sundrum models \cite{Randall:1999ee}. The signals under study are non-resonant $HH$ production with 
event kinematics predicted by the SM and resonant $HH$ production 
with event kinematics consistent with the decays of heavy spin-0 or spin-2
resonances.

Previous searches for $pp \to HH$ production were performed 
by the ATLAS and CMS collaborations in Run 1 of the LHC at  
$\sqrt{s} =8$~\TeV. Decay modes with $4b$ \cite{EXOT-2014-11, CMS-HIG-14-013}, 
$b \bar{b} \tautau$ \cite{HIGG-2013-33, CMS-HIG-14-034},  
$\gamma \gamma b \bar{b}$ \cite{CMS-HIG-13-032, HIGG-2013-29} 
and $\gamma \gamma WW^{\ast}$ \cite{HIGG-2013-33} in the final state were 
studied. Furthermore, ATLAS also published a combination of all of 
the explored channels~\cite{HIGG-2013-33}. 

Results at $\sqrt{s} =13$~\TeV\ 
were published by the ATLAS Collaboration in the $4b$ \cite{EXOT-2016-31}, $b \bar{b}\tautau$\cite{bbtautau},
$b \bar{b}\gamma \gamma$\cite{bbgammagamma} and
$WW\gamma\gamma$\cite{WWgammagamma} 
decay mode and by CMS in the $4b$ \cite{Sirunyan:2018zkk} , $b \bar{b}\tautau$
\cite{CMS-HIG-17-002}, $b \bar{b}\gamma\gamma$ \cite{CMS-HIG-17-008} and in the $b\bar{b}WW^{\ast}$ channel using 
the dileptonic $WW^{\ast}$ decay mode \cite{CMS-HIG-17-006}.  
Given the low expected yield for SM $HH$ non-resonant production, 
it is of great importance to understand the sensitivity for the observation of the Higgs boson pair production in all possible decay channels, including $b \bar{b}WW^{\ast}$,
which will improve projections for future high-luminosity 
and high-energy colliders.

\begin{figure}[!h]
\begin{center}
\includegraphics*[width=0.35\textwidth] {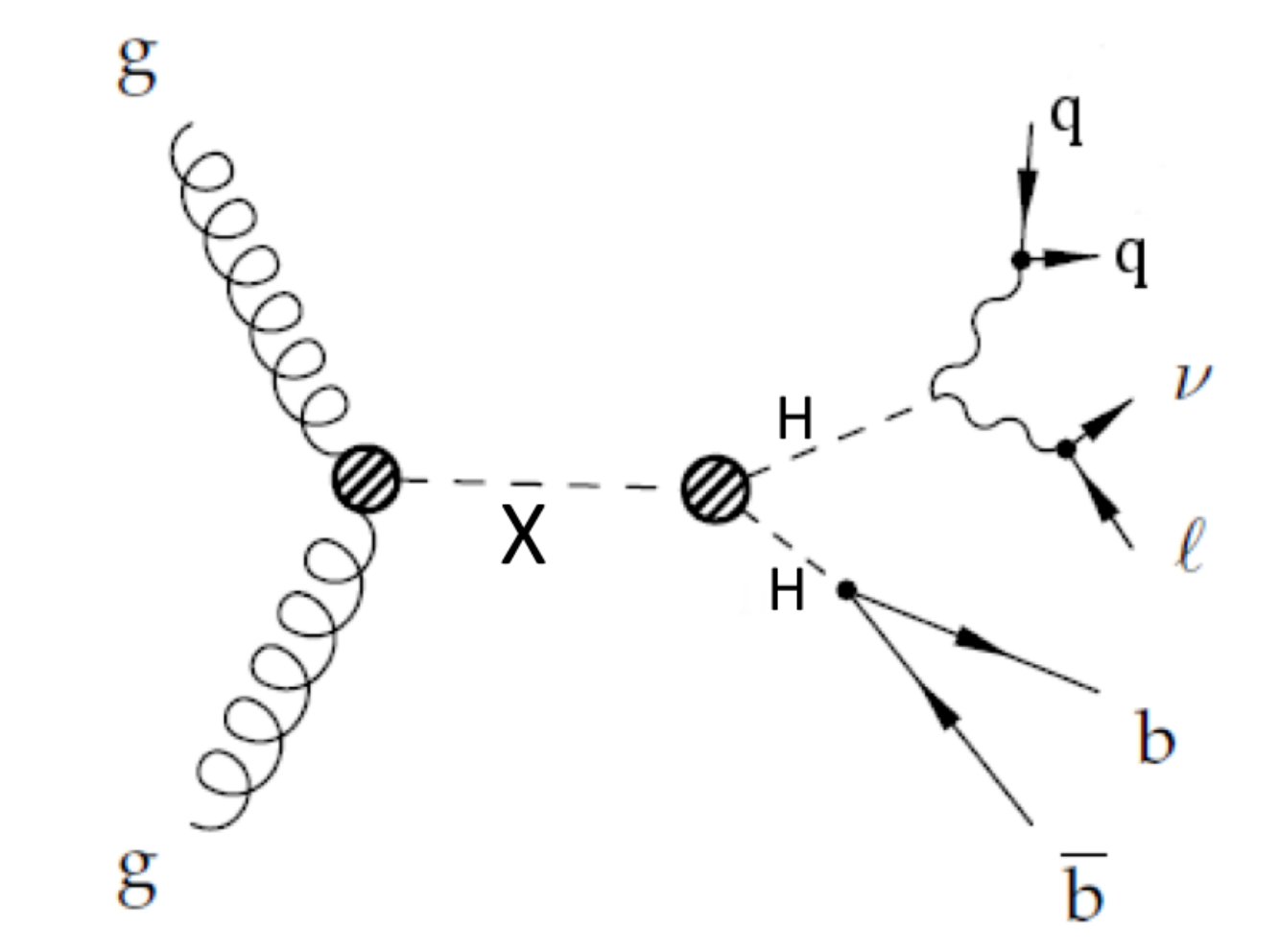}
  \caption{Schematic diagram of resonant Higgs boson pair production 
    with the subsequent Higgs and $W$ boson decays.}
  \label{fig:resonant_hh_production}
\end{center}
\end{figure}

This paper reports results of a search for Higgs boson 
pair production where one Higgs boson decays via $H \to b \bar{b}$, 
and the other decays via $H \to WW^{\ast}$. The $H \to WW^{\ast}$ branching fraction is the second largest 
after $H \to b \bar{b}$, so the $ b \bar{b}WW^{\ast}$ final state can
be sensitive to $HH$ production if the signal can be well separated
from the dominant $t\bar{t}$ background.
The $WW^{\ast}$ system 
decays into  $\ell\nu qq$ (where $\ell$ is either an electron or a
muon), and the small contamination from leptonic $\tau$ decays is not
explicitly vetoed in the analysis.
Figure~\ref{fig:resonant_hh_production} shows 
a schematic diagram of resonant production of the Higgs boson pair 
with the subsequent decays $H \to WW^{\ast}$ 
and $H \to b \bar{b}$. 

Two complementary techniques are used to reconstruct the Higgs
boson candidate that decays into two $b$-quarks. Both techniques use the
anti-$k_t$ jet algorithm~\cite{Cacciari:2008gp} but with different radius
parameters. The first technique employs jets with  radius parameter 
$R = 0.4$ and it is used when  each $b$-quark from the $H\rightarrow b \bar{b}$ decay can be
reconstructed as a distinct $b$-jet.  The second technique is used
when this is not possible, due to the large boost of the $b$-quark pair.
In this case the Higgs boson candidate is identified as a single anti-$k_t$ jet with radius
parameter $R = 1.0$. The analysis using the first technique is 
referred to as the ``resolved'' analysis and that using the second
technique is referred to as the ``boosted'' analysis. In both analyses, the jets from the hadronically decaying $W$ boson are reconstructed as anti-$k_t$ jets with radius parameter $R = 0.4$. 
The resonant $HH$ search is performed using either the resolved or boosted analysis method depending on which is the most sensitve to the particular model and $HH$ mass being tested, in contrast to the non-resonant search which uses only the resolved analysis method. 

The dominant background in the $b \bar{b} WW^{\ast}$ final state is $t \bar{t}$ production, with smaller 
contributions from $W$ bosons produced in association with jets ($W$+jets) and  multijet events in which a jet is
misidentified as a lepton. 
The analysis defines one signal region for each signal hypothesis and,
in order to avoid biases in the analysis selection, 
the analysis procedures and the event selection were optimised without reference 
to data in the signal regions.

\section{Data and simulation samples}
\label{sec:samples}
 The ATLAS detector~\cite{PERF-2007-01} is a general-purpose particle detector at the Large Hadron Collider optimised to discover and measure a broad range of physics processes.  It consists of
an inner tracking detector surrounded by a thin superconducting
solenoid, electromagnetic and hadronic calorimeters, and a muon
spectrometer incorporating three large superconducting toroid magnets.\footnote{ATLAS uses a right-handed coordinate system with its origin at the nominal interaction point (IP) in the centre of the detector and the
  $z$-axis along the beam pipe. The $x$-axis points from the IP to the
  centre of the LHC ring, and the $y$-axis points upwards.
  Cylindrical coordinates $(r,\phi)$ are used in the transverse plane,
  $\phi$ being the azimuthal angle around the $z$-axis. The
  pseudorapidity is defined in terms of the polar angle $\theta$ as
  $\eta = -\ln \tan(\theta/2)$. The angular distance is measured in units
  of $\Delta R \equiv \sqrt{(\Delta\eta)^{2} + (\Delta\phi)^{2}}$.}

The dataset used in this analysis corresponds to an integrated 
luminosity of 36.1~fb$^{-1}$ (3.2~fb$^{-1}$ from 2015 and 
32.9~fb$^{-1}$ from 2016) recorded by  single-electron or 
single-muon triggers. The single-lepton trigger efficiency ranges from 75\% to 90\% (75\% to 80\%) for electrons (muons) depending on the signal mass, 
for selected lepton candidates above \pt thresholds defined in
Section \ref{sec:resolved_selection}. 
Samples of simulated signal and background events were used to
design the event selection and estimate the signal acceptance and
the background yields from various SM processes. 

When searching for a new resonance (denoted by $X$ in the following), specific simulation models must be employed.
Therefore, the spin-0 states were treated as narrow heavy neutral Higgs bosons, while the
spin-2 states were modelled as narrow Randall--Sundrum (RS) gravitons
\cite{Agashe:2007zd, Fitzpatrick:2007qr}.
The parameters used in the RS graviton simulation were: 
 $c = k/\bar{M}_{\rm Pl}$ equal to 1.0 or 2.0, where $k$ is the 
curvature of the warped extra dimension and $\bar{M}_{\rm Pl}=2.4\times 10^{18}$~\gev\
is the effective four-dimensional Planck scale.
The graviton signal samples were generated at leading order (LO) with
\textsc{Madgraph5\_aMC@NLO} \cite{Alwall2014}  using the NNPDF2.3~\cite{Ball:2012cx} LO parton distribution function  (PDF) set, 
and \textsc{Pythia}~8.186~\cite{Sjostrand:2008vc} to model
the parton showers and hadronisation process with a set of tuned underlying-event parameters 
called the A14 tune~\cite{ATL-PHYS-PUB-2014-021}. 
Only the $c=2.0$ samples were fully simulated, while the $c=1.0$ samples were
obtained by reweighting them using the Monte Carlo (MC) generator-level $m_{HH}$ distribution.

Scalar signal samples were generated at next-to-leading order (NLO) 
with \textsc{Madgraph5\_aMC@NLO} 
interfaced to \textsc{Herwig++}~\cite{Bahr:2008pv} using the CT10 
PDF set \cite{Lai:2010vv} and the UE-EE-5-CTEQ6L1 tune. 
The simulation
produced the Higgs boson pair through gluon--gluon fusion using an effective field
theory approach to take into account the finite value of the top-quark mass $m_{t}$~\cite{Hespel:2014sla}. 
Events were first generated with an effective 
Lagrangian in the infinite top-quark mass approximation, and then 
reweighted  with form factors that take
into account the finite mass of the top quark. 

The non-resonant signal samples were
simulated with  \textsc{Madgraph5\_aMC@NLO} + \textsc{Herwig++} using the CT10 
PDF set; and the same approach for the inclusion of finite  $m_{t}$
effects was used \cite{Frederix:2014hta}. In addition, scale factors dependent on the $HH$ invariant mass $m_{HH}$ at  generator level were applied to match the MC $m_{HH}$ distribution with an NLO calculation that computes exact finite $m_{t}$
contributions~\cite{Borowka:2016ehy}. All signal samples were generated with 100\% of Higgs boson pairs decaying into $ b \bar{b}WW^{\ast}$, and the samples were then normalised assuming ${\mathcal{B}}(H \to WW^*) = 0.22$ and ${\mathcal{B}}(H \to bb) = 0.57$ \cite{deFlorian:2016spz}.

\textsc{Sherpa} v2.2~\cite{Gleisberg:2008ta} with the 
\textsc{NNPDF 3.0}~\cite{Ball:2014uwa} PDF set was used as the baseline generator for the 
($W \to \ell\nu$)/($Z\to \ell\ell$)+jets background. 
The $W/Z$+jets samples were normalised using the FEWZ~\cite{Gavin:2010az}
inclusive cross section with  NNLO accuracy. The diboson 
processes ($WW$, $WZ$ and $ZZ$) were generated at NLO  with \textsc{Sherpa} v2.1.1~\cite{Gleisberg:2008ta}
with the \textsc{CT10}~\cite{Lai:2010vv} PDF set and normalised using
the \textsc{Sherpa} cross-section prediction.

The $t\bar{t}$ background samples were generated with
\textsc{Powheg-Box} v2~\cite{Frixione:2007vw} using the \textsc{CT10} PDF set. \textsc{Powheg-Box} v2 was interfaced to \textsc{Pythia} 6.428~\cite{Sjostrand:2006za} for parton 
showers, using the \textsc{Perugia2012}~\cite{Skands:2010ak} tune with the 
CTEQ6L1~\cite{Pumplin:2002vw} set of PDFs for the underlying-event description.
\textsc{EvtGen} v1.2.0~\cite{Lange:2001uf} was used to simulate the
bottom and charm hadron decays. The mass of the top quark was set to
$m_{t} = 172.5~\GeV$. At least one top quark in the $t\bar{t}$ event was
required to decay into a final state with a lepton. For the $t\bar{t}$
sample the parameter \textsc{Hdamp},
used to regulate the high-\pt\ gluon emission in \textsc{Powheg}, was set
to $m_{t}$, giving good modelling of the high-\pt\ 
region~\cite{ATL-PHYS-PUB-2014-005}. The interference between the
$t\bar{t}$ background and the signal is extremely small due to the
 small width of the Higgs boson ($\Gamma_H\sim$4 MeV) and it
has been neglected in this analysis.  The $t\bar{t}$ 
cross section is calculated to next-to-next-to-leading order in QCD including resummation of soft gluon
contributions at next-to-next-to-leading-logarithm (NNLL) accuracy using
\textsc{Top++} 2.0~\cite{Czakon:2011xx}.  

Single-top-quark events in the $Wt$-, s- and t-channels were generated using \textsc{Powheg-Box} v1~\cite{Alioli:2009je, Re:2010bp}. 
The overall normalisation of single-top-quark production in
each channel was rescaled according to its 
approximate  NNLO cross section~\cite{Kidonakis:2011wy, Kidonakis:2010ux, Kidonakis:2010tc}.  

The effect of multiple $pp$ interactions in the same and neighbouring 
bunch crossings (pile-up) was included by overlaying minimum-bias 
collisions, simulated with {\sc Pythia}\,8.186, 
on each generated signal and background event. The interval between
proton bunches was 25~ns in all of the data analysed. The number of overlaid 
collisions was such that the distribution of the number of interactions 
per $pp$ bunch crossing in the simulation matches that observed in 
the data: on average 14 interactions per bunch crossing in 2015 and 
23.5 interactions per bunch crossing in 2016. The generated
samples were processed through a \textsc{Geant}4-based detector 
simulation~\cite{Agostinelli:2002hh, SOFT-2010-01} with the
standard ATLAS reconstruction software used for collision data. 

\section{Object reconstruction}
\label{sec:objects}
In the present work an ``object'' is defined to be a reconstructed jet, electron, or muon.
Electrons are required to pass the ``TightLH'' selection as described in Ref.~\cite{Aaboud:2016vfy, ATLAS-CONF-2016-024}, have $\pt > 27$~\gev and be within
$|\eta|<2.47$, excluding the transition region between the barrel and
endcaps in the LAr calorimeter ($1.37 < |\eta| < 1.52$). In addition,
the electron is required to be isolated. In order to calculate the
isolation variable, the $\pt$ of the tracks in a cone of 
$\Delta R$ around the lepton track is summed ($\sum \pt$), where 
$\Delta R = {\mathrm{min}}(10~\GeV/{\pt^{e}}, 0.2)$ and $\pt^{e}$ is 
the electron transverse momentum. The ratio $\sum \pt/\pt^{e}$
(isolation variable) is required to be less than 0.06.

Muons are reconstructed as described in Ref.~\cite{Aad:2016jkr} and required to pass
the ``Medium'' identification criterion and have $|\eta|<2.5$. The muon isolation
variables are similar to the electron isolation variables with the
only difference being that the maximum cone size is $\Delta R = 0.3$ 
rather than 0.2.

Jets are reconstructed 
using the anti-$k_t$ algorithm~\cite{Cacciari:2008gp} with a radius parameter of 0.4, and are required to have $\pt > 20$~\gev ~and $|\eta|<2.5$. 
Suppression of jets likely to have originated from pile-up interactions is achieved using  
a boosted decision tree in an
algorithm that has an efficiency of 90\% for jets with $\pt < 50$~\GeV\ and $|\eta|< 2.5$~\cite{ATLAS-CONF-2014-018}.

Large-$R$ jets are reconstructed using the anti-$k_t$ algorithm with a radius parameter of 1.0 
and are trimmed to reduce pile-up contributions to the jet, as described in Ref.~\cite{Krohn:2009:Trimming}. 
The jet mass ($m_\text{J}$) resolution is improved at 
high momentum by using tracking in addition to calorimeter information~\cite{ATLAS-CONF-2016-035}.
Large-$R$ jets are required to have \pt > 250 \GeV, $m_\text{J}$ > 30 \GeV\ and $|\eta|< 2.0$. The identification of large-$R$ jets consistent with boosted Higgs boson decays uses
jets built from ID tracks (referred to as track-jets) to identify the $b$-jets 
within the large-$R$ jets. The track-jets are built with the anti-$k_t$ algorithm with $R = 0.2$~\cite{ATLAS-CONF-2016-039}. They are required to have \pt > 10 \GeV, $|\eta|< 2.5$, and are matched to 
the large-$R$ jets with a  ghost-association algorithm~\cite{Cacciari:2008:Ghost}.

The jet-flavour tagging algorithm~\cite{Aaboud:2018xwy} is used to
select signal events and to suppress multijet, $W$+jets, $Z$+jets and
diboson backgrounds. The jets containing  $b$-hadrons are called
$b$-jets in this work. The jet-flavour tagging algorithm parameters
were chosen such that the $b$-tagging efficiency is 85\% for jets with \pt of at least 20~\gev~as
determined in simulated inclusive \ttbar events~\cite{Aaboud:2018xwy} . At this efficiency,
for jets with a $\pt$ distribution similar to that originating from
jets in \ttbar events, the charm-quark component is suppressed by a factor 
of 3.1 while the light-quark component is suppressed by a factor of
34.  Jets that are not tagged as $b$-jets are collectively referred to as ``light jets". 

The calorimeter-based missing transverse momentum with magnitude
$\met$ is calculated as the negative vectorial sum of the 
transverse momenta of all calibrated selected objects, such as 
electrons and jets, and is corrected to take into account the transverse
momentum of muons. 
Tracks with 
$\pT^{\mathrm{track}} > 500~\MeV$, compatible with the primary 
vertex but not matched to any reconstructed object, are included
in the reconstruction to take into account the soft-radiation
component that does not get clustered into any hard 
object~\cite{ATL-PHYS-PUB-2015-023}.
                                         
To avoid double-counting, overlapping objects are removed from the analysis according to the
following procedure. Muons sharing their track with an electron  are
removed if they are calorimeter-tagged. Otherwise, the electron is removed.
 Jets overlapping with electrons within an angular distance $\Delta R = 0.2$ are removed. Jets overlapping with muons within $\Delta R = 0.2$ and having less than three tracks or carrying less than 50\%  of 
the muon $\pt$ are removed.  Electrons overlapping with remaining
jets within $\Delta R = {\mathrm{min}}(0.4, 0.04 + 10~\GeV{}/\pt^{e})$ are removed. Muons overlapping with
remaining jets within $\Delta R  = {\mathrm{min}}(0.4, 0.04 + 10~\GeV{}/\pt^{\mu})$ are removed.
s

\section{Resolved analysis}
\subsection{Resolved analysis: event selection}
\label{sec:resolved_selection}
At lowest order in QCD the final-state particles consist of one charged lepton, one
neutrino, and jets of colourless hadrons from four quarks, two being $b$-quarks. Therefore, the corresponding detector
signature is one charged lepton ($e$/$\mu$), large $\met$,
and four or more jets. Two of these jets are $b$-tagged jets from the Higgs boson decay, 
and two jets are non-$b$-tagged jets from the hadronic $W$ boson decay. 

The data used in the analysis were recorded by several single-electron or single-muon triggers in 2015 and 2016. In 2015, the electron (muon) trigger required a $p_{\rm T} > 24~(20) $ GeV electron (muon) candidate. Because of a higher instantaneous luminosity, in 2016 the electron trigger required a $p_{\rm T} > 26$ GeV electron candidate, while muons were triggered using a \pt threshold of 24 GeV at the beginning of data taking, and 26 GeV for the rest of the year. In both 2015 and 2016, a threshold of  $p_{\rm T} > 27$ GeV was applied offline on the selected lepton candidate.

The analysis selects events that contain at least one reconstructed
electron or muon matching a trigger lepton candidate. 
In order to ensure that the leptons originate from the interaction 
point, requirements on the transverse ($d_0$) and longitudinal ($z_0$)
impact parameters of the leptons relative to the primary vertex
are imposed. In particular, defining  $\sigma_{d_0}$ as the uncertainty in the measured 
$d_0$ and $\theta$ as the angle of the track relative to the beam
axis,  the requirements $|d_0|/\sigma_{d_0}<2$  and
$|z_0\sin\theta|<0.5$~mm are applied.  The 
requirement on $|d_0|/\sigma_{d_0}$ is relaxed to define control regions 
in order to estimate the multijet background. 
The highest $p_T$ lepton is then retained as the analysis lepton.

Events are required to have exactly two 
$b$-tagged jets, which form the Higgs boson candidate. 
Since events are accepted if they contain two or 
more light jets, in events with more than two light jets, the three 
leading jets are considered, and the pair with the lowest $\Delta R$ 
between them is selected as the $W$ boson candidate. From MC simulation it
was found that, 
when the light quarks from the $W$ boson are matched to reconstructed jets by requiring that
the $\Delta R$ between the jet
and the quark is less than 0.3,  this procedure yields 
the correct jet assignment in \makebox{70\%} of the cases.

The event kinematics of the 
$H \to WW^* \to \ell\nu qq$ topology can be fully reconstructed. 
Among all four-momenta of the final-state particles, only 
the component of the neutrino momentum along the beam axis, referred to as longitudinal momentum ($p_z$) in the
following, is unknown  
while its transverse momentum is assumed to be  the $\met$. The neutrino longitudinal
momentum is computed by solving a quadratic 
equation in $p_z$, employing the four-momenta of the lepton and the
hadronic $W$ boson, the $\met$, and the $m_{H} = 125$ GeV constraint on the $WW^*$ system. 
No $W$ boson mass constraint is applied to either the hadronic or the leptonic $W$ boson decay, allowing either $W$ boson to be off-shell.  Whenever two real solutions are obtained, 
the $\nu$ candidate with the smallest $\Delta R$ relative to the lepton 
direction is retained. Studies performed by matching the $\nu$ candidate with  the MC
generator-level neutrino show that this procedure finds the
correct solution for the neutrino $p_z$ in 60\% (75\%) of cases for a resonant signal of mass 700 (3000) \gev. If two complex solutions are found, only 
the real part of the solutions is retained. With the
neutrino longitudinal momentum computed, the di-Higgs invariant mass can be fully reconstructed and employed to discriminate against backgrounds. 

Kinematic selections are used to suppress the $t\bar{t}$ 
background relative to the signal. The $t \bar{t}$ events are 
typically characterised by two $b$-jets and two $W$ bosons such 
that the $\Delta R$ separation between the $b$-jets is large, and
similarly the $\Delta R$ separation between the $W$ bosons is also
large. In contrast, in particular when the invariant
mass of the heavy resonance is large, the signal is characterised by two
$b$-jets and two $W$ bosons which are closer in $\Delta R$ in signal
events with respect to the $t \bar{t}$ background events.  
Moreover, for the signal
the two $b$-jets have an invariant mass equal to $m_H$, while this is not
the case for the $t\bar{t}$ background, where a much broader
distribution is expected. The symbols of the kinematic variables that
discriminate between signal and background are listed in Table~\ref{tab:kin_var}.
\begin{table}
\caption{Selection variables used to identify the $HH \to b\bar{b}WW^{\ast}$ decay
  chain in the resolved analysis. The $m_{WW^{\ast}}$ variable is exactly  equal to $m_H$ if a
  real solution for the neutrino $p_z$ is found. It is larger
  otherwise.} \label{tab:kin_var}
\begin{center}
\begin{tabular}{l|l}
\multicolumn{2}{c}{Definition of the $HH \to b\bar{b}WW^{\ast}$ kinematic variables} \\
\hline
 $p_{\rm T}$ of the $b \bar{b}$ pair  & $p_{\rm T}^{b \bar{b}}$ \\
 $p_{\rm T}$ of the $WW^{\ast}$ pair         & $p_{\rm T}^{WW^{\ast}}$ \\
 $\Delta R$ of the $WW^{\ast}$ pair         & $\Delta R_{WW^{\ast}}$ \\
 $WW^{\ast}$ pair mass                      & $m_{WW^{\ast}}$ \\
 $b \bar{b}$ pair mass & $m_{bb}$ \\
Di-Higgs boson system invariant mass & $m_{HH}$ \\
\end{tabular}
\end{center}
\end{table}

The selection requirements on the kinematic variables defining the signal region  were chosen
to maximise the expected sensitivity to various signals.
The optimisation  was performed for a spin-0 signal considering
resonance masses  ($m_X$)  from 500~\GeV\ to 3000~\GeV\ in steps of 100~\GeV.  The same selection was used
for the spin-2 signal models while SM Higgs pair production was 
used to optimise the non-resonant analysis. Below 500~\GeV\ the top-quark background increases significantly, and hence rapidly reduces sensitivity.

The selection criteria define four sets of requirements, referred as 
\emph{non-res}, \emph{m500}, \emph{low-mass} and \emph{high-mass} in
the following. They are 
shown in Table~\ref{tab:sig_reg_summary}. The \emph{non-res} and \emph{m500} selections are exclusively used for non-resonant signal and resonant signal with mass 500~\GeV\ respectively.
The \emph{low-mass} selection is used for signal masses from 600 to
1300~\GeV, while the \emph{high-mass} selection is used for signals with
masses between 1400 and 3000~\GeV. 
In addition, requirements are placed on the reconstructed
di-Higgs invariant mass $m_{HH}$ as a function of the signal 
resonance mass $m_{X}$, as shown in Table \ref{tab:mhh_cuts}. The
resolution of the reconstructed $m_{HH}$ ranges from 6\% at 500~\GeV\ to
10\% at 3000~\GeV.

\begin{table}
\caption{Criteria for non-resonant, \emph{m500}, \emph{low-mass} and 
    \emph{high-mass} selections in the resolved analysis. 
    } 
\label{tab:sig_reg_summary}
\begin{center}
\begin{tabular}{l|c|c|c|c}
 Variable & \emph{non-res} & \emph{m500} & \emph{low-mass} & \emph{high-mass}\\
\hline
 $\met$ [GeV]		&$> 25$&$>25$&$> 25$& $> 25$ \\
$m_{WW^*}$ [GeV] 	   		&$< 130$ &$< 130$ & $< 130$& none\\
$p_{\rm T}^{b \bar{b}}$ [GeV] 		&$> 300$& $> 210$&$> 210$&$> 350$\\
$p_{\rm T}^{WW^*}$ [GeV] 		&$> 250$&$> 150$&$> 250$&$> 250$ \\
$\Delta R_{WW^*}$  		& none & none & none &$<1.5$\\
$m_{b \bar{b}}$ [GeV]   		&105--135&105--135&105--135&105--135\\
\end{tabular}
\end{center}
\end{table}

\begin{table}
\caption{Window requirements on $m_{HH}$ as a function of the resonance mass
  $m_{X}$ in the resolved analysis.}
\label{tab:mhh_cuts}
\begin{center}
\begin{tabular}{l|r|r|r|r|r}
$m_{X}$ [GeV]      &   500   &   600   &   700   &   750   &   800\\ 
$m_{HH}$ window [GeV] & 480--530 & 560--640 & 625--775 & 660--840 & 695--905\\

\hline 
$m_{X}$ [GeV]        & 900           & 1000 	    &  1100   &  1200  & 1300 \\  
$m_{HH}$ window [GeV]  & 760--967   & 840--1160 & 925--1275	&1010--1390 &1095--1505  \\

\hline 
$m_{X}$ [GeV]       & 1400 &  1500 &  1600  & 1800   & 2000\\
$m_{HH}$ window [GeV] & 1250--1550 &  1340--1660  & 1430--1770 &  1750--2020 & 1910--2170\\

\hline 
 $m_{X}$ [GeV]           & 2250           &  2500        &  2750   		& 3000  \\
$m_{HH}$ window [GeV] 	    &  2040--2460 & 2330--2740 &  2570--2950	& 2760--3210 \\

\end{tabular}
\end{center}
\end{table}
  
\subsection{Resolved analysis: background determination}
\label{sec:resolved_backgrounds}
In this analysis the presence of a signal is indicated by an excess of events 
over the SM prediction for the background yield in the signal regions,
so it is of great importance to properly estimate the amount of
background in those regions.
The dominant background is the $t\bar{t}$ process.
Dedicated control regions are used to normalise and validate the estimate of this background.
The $t\bar{t}$ normalisation is performed using three data
control regions, one for the \emph{non-res},
a second one for the \emph{m500} and \emph{low-mass},  and a third one for 
the  \emph{high-mass} selection. These control regions are obtained
by selecting events outside the $m_{b \bar{b}}$ window [100, 140]~\GeV\ and
applying only the $\met$,  $m_{WW^*}$ (where applicable) and $p_{\rm T}^{b \bar{b}}$ requirements shown in 
Table~\ref{tab:sig_reg_summary} for the respective selections.

In all regions, the event yields of $W/Z$+jets, 
single-top-quark and diboson events are modelled using
simulated events and normalised to the expected SM cross sections. 

The multijet component of the background
originates from events where either a jet is incorrectly identified 
as a lepton, or a non-prompt lepton is produced in heavy-flavour
decays, or from photon conversions. It is characterised by  low
$\met$ and  high $|d_0|/\sigma_{d_0}$ values of the lepton.  The multijet background makes a significant contamination in the top control regions. Therefore, this background is estimated in each top control region and signal region using a data-driven
two-dimensional sideband method, labelled the ABCD method, that uses three additional regions
 denoted in the following by  B, C and D. The region of
interest, signal or control region, is indicated by A.

The  B, C and D regions are defined in the following way:
\begin{itemize}
\item region B:  $\met < 25~\GeV$ and  $|d_0|/\sigma_{d_0}<2.0$,
\item region C: $\met > 25~\GeV$ and $|d_0|/\sigma_{d_0}>2.0$, and 
\item region D:  $\met < 25~\GeV$ and  $|d_0|/\sigma_{d_0}>2.0$. 
\end{itemize}
while $N_A$,$N_B$,$N_C$ and $N_D$ indicate the number of events in the
$A$,$B$,$C$ and $D$ regions respectively.
In the absence of correlations between the $\met$ and $|d_0|/\sigma_{d_0}$ variables, the relation
$N_{\text{A}} = N_{\text{C}} N_{\text{B}}/N_{\text{D}}$ holds, while in practice a correlation
among variables results in a correction factor $F$ to be 
applied to the computed ratio $N_{\text{A}}^{\text{corrected}} = FN_{\text{C}} N_{\text{B}}/N_{\text{D}}$.
The correction factor $F$ is estimated from data at an early stage of
the analysis selection once a veto on the signal candidates is applied
by inverting the requirement on the $m_{b \bar{b}}$ variable. It is
computed using the relation $F= N_{\text{A}} N_{\text{D}}/(N_{\text{C}} N_{\text{B}})$.
 Systematic uncertainties in $F$ are described in Section~\ref{sec:systematics}.
In order to reduce statistical uncertainties in the computation,  the shape of the $m_{b \bar{b}}$ distribution is derived at an earlier stage 
of the selection sequence, after applying  the $m_{WW^*} < 130~\GeV$ and
$\pt^{b \bar{b}} > 210~\GeV$  requirements for the \emph{non-res}, \emph{m500} and \emph{low-mass} analyses 
and the $\pt^{b \bar{b}} > 350~\GeV$ and $\pt^{WW^*} > 250~\GeV$ requirements for 
the \emph{high-mass} analysis. 
It was verified that subsequent requirements do not affect the 
$m_{b \bar{b}}$ shape, which can therefore be used at the end of the selection sequence. 
Table~\ref{tab:CRs} summarises the numbers of observed  and 
estimated events in the three top-quark control regions.  The event yields in the control regions are
used as input to the statistical analysis. Major contamination in the $t
\bar{t}$ control regions comes from multijet and $W$+jets backgrounds; as a result
the $t \bar{t}$ purity ranges from 52\% to 58\%.

\begin{table}
\caption{Data and estimated background yields in the \emph{non-res}, 
\emph{m500} and \emph{low-mass}, and \emph{high-mass} top-background control regions of the resolved analysis. 
The uncertainty shown for the multijet background is due to the number of data events
in the C region. For all other backgrounds the uncertainties are due to the 
finite MC sample sizes.}
\label{tab:CRs}
\begin{center}
\begin{tabular}{l*{3}{r@{}c@{}l}}
  Process & \multicolumn{3}{c}{\emph{non-res}} &
                                                 \multicolumn{3}{c}{\emph{m500}~and~\emph{low-mass}} & \multicolumn{3}{c}{\emph{high-mass}}\\

\hline
\ttbar				&110 &$\,\pm\,$& 6 & ~~~~~~~~~532&$\,\pm\,$&13& 8570&$\,\pm\,$&50\\
Multijet	 	        &  33&$\,\pm\,$&4 & 250&$\,\pm\,$&30& 1540&$\,\pm\,$&250 \\
$W$+jets	 	        &  29&$\,\pm\,$&1 & 125&$\,\pm\,$&3& 2259&$\,\pm\,$&8 \\
Single top		        & 20&$\,\pm\,$&2	& 76&$\,\pm\,$&4	  & 1780&$\,\pm\,$&20 \\
Dibosons			& 2.2&$\,\pm\,$&0.4  &  8.3&$\,\pm\,$&0.8	& 171&$\,\pm\,$&4\\
$Z$+jets	  		& 6.7&$\,\pm\,$&0.2& 27.1&$\,\pm\,$&0.8	 & 404&$\,\pm\,$&2 \\
Background sum	& 201&$\,\pm\,$&8 & 1015&$\,\pm\,$&34& 14720&$\,\pm\,$&260\\
\hline
Data				& \multicolumn{3}{c}{206}&  \multicolumn{3}{c}{1069} & \multicolumn{3}{c}{14862} \\
\end{tabular}
\end{center}
\end{table}

\begin{figure}
\includegraphics*[width=0.5\textwidth]{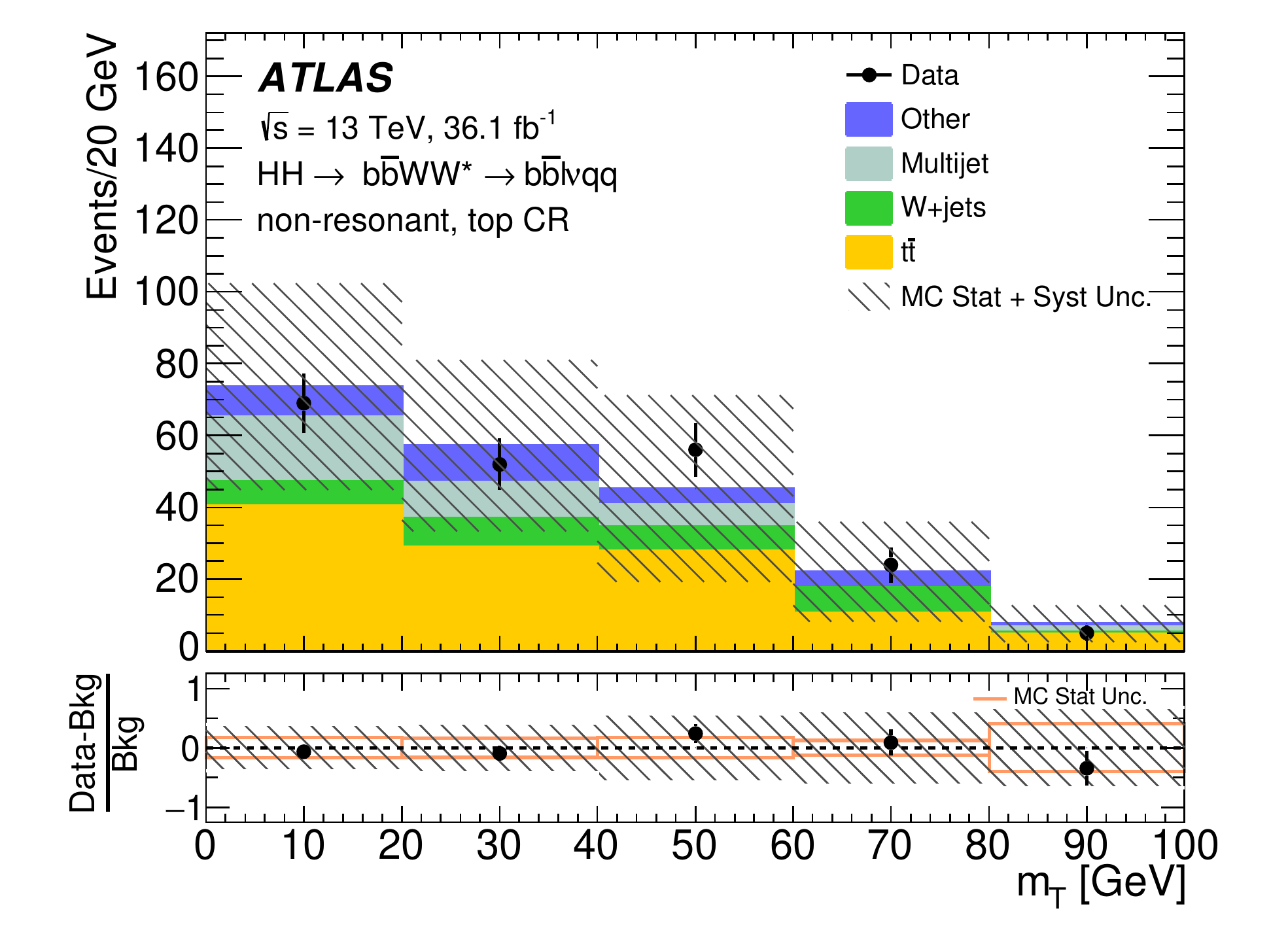}
\includegraphics*[width=0.5\textwidth]{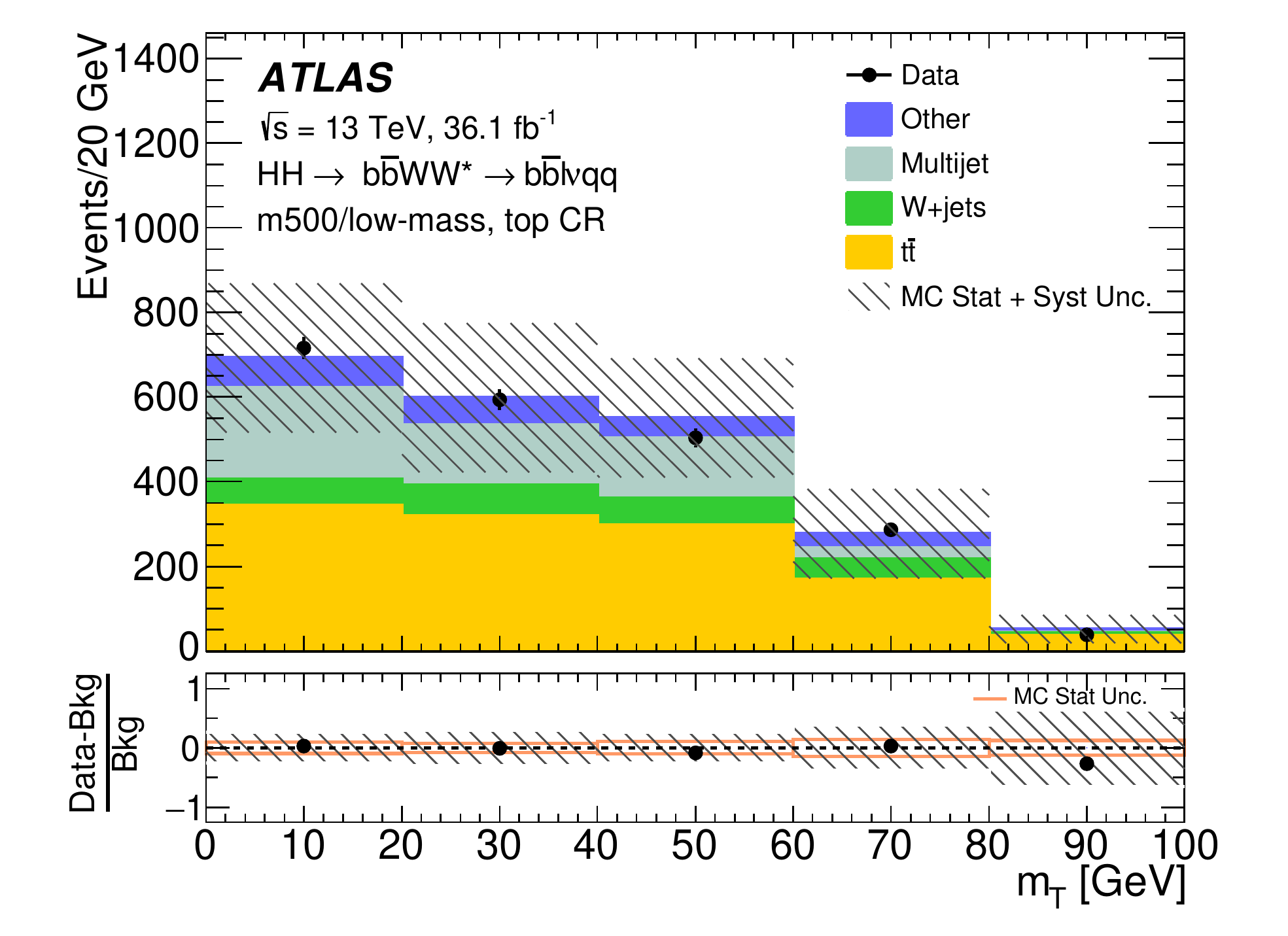}
\includegraphics*[width=0.5\textwidth]{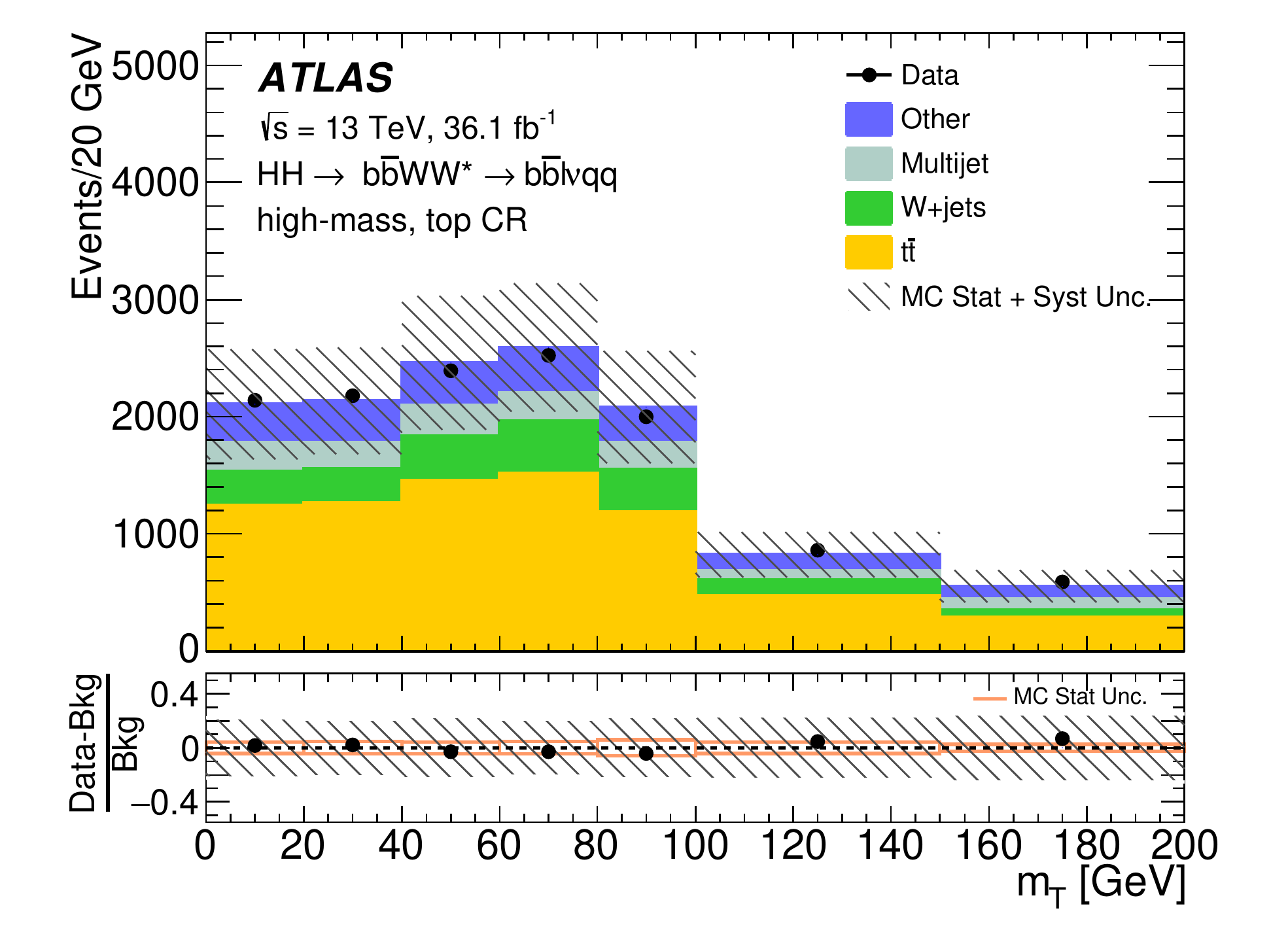}
\caption{The $\mT$ distribution in the three top-background control regions for the \emph{non-res}, \emph{low-mass}, and 
  the \emph{high-mass} selections of the resolved analyses. 
  The signal contamination is negligible, and hence not shown. The lower panel shows the 
  fractional difference between the data and the total expected background 
  with the corresponding statistical and total uncertainty. } 
  \label{fig:resolved_mtlep}
\end{figure}

 The modelling of the background was checked at all selection 
stages and, in general, shows good agreement with data.
Figure~\ref{fig:resolved_mtlep} shows the $\mT$ distribution of the
leptonic $W$ boson candidate  in the three 
top control regions. The $\mT$ variable is defined as:
\[
\mT = \sqrt{2\pt^{\ell}\met\cdot (1-\rm{cos}\Delta \phi)} ~,
\]
where $\Delta \phi$ is the azimuthal 
angle between $\pt^{\ell}$ and $\met$. The multijet background populates the 
low values of the $\mT$ distribution, so any mis-modelling of the multijet background would be clearly visible in the $\mT$ distribution. 

\begin{figure}
\begin{flushleft}
\includegraphics*[width=0.45\textwidth]{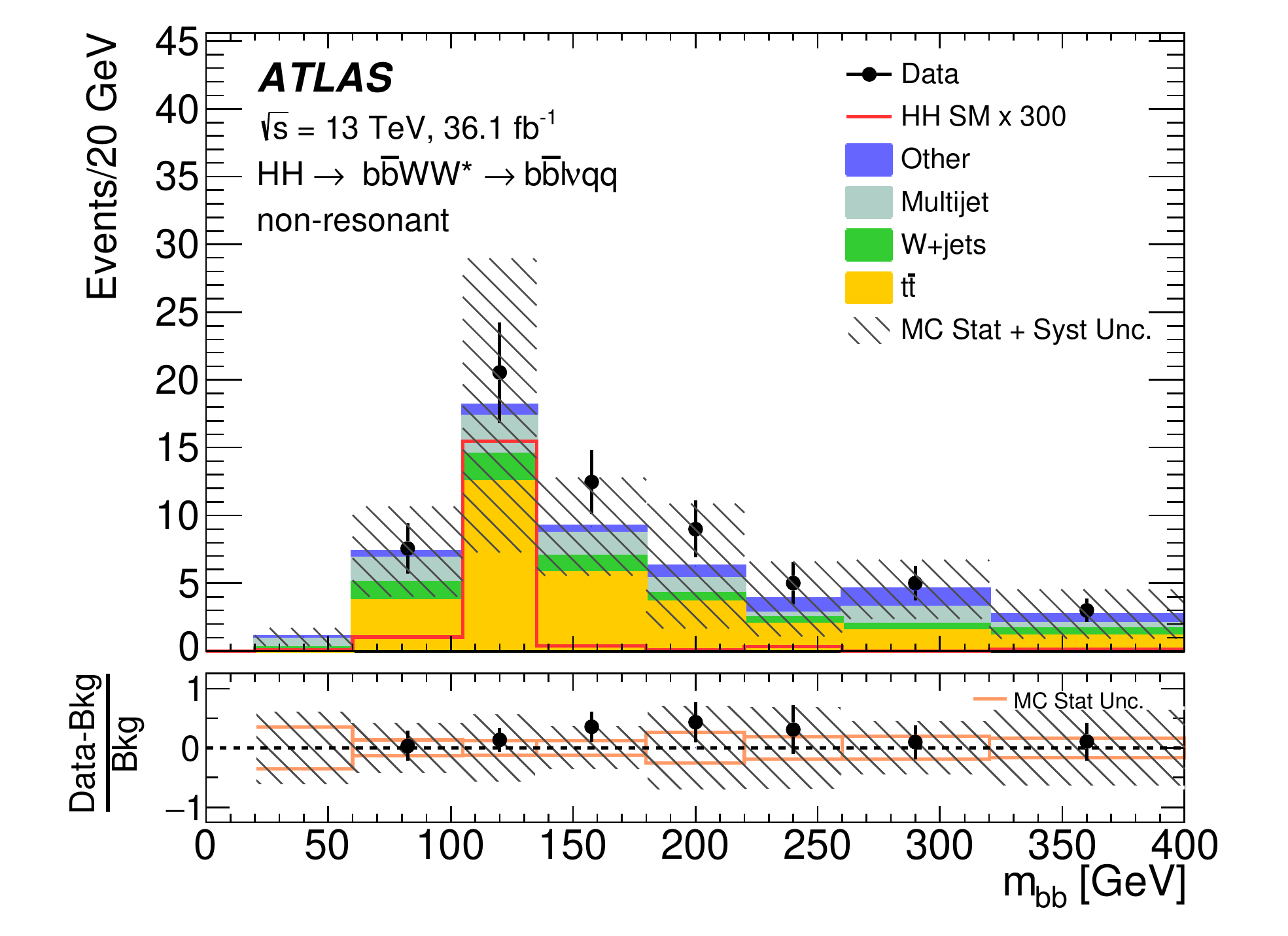}
\includegraphics*[width=0.45\textwidth]{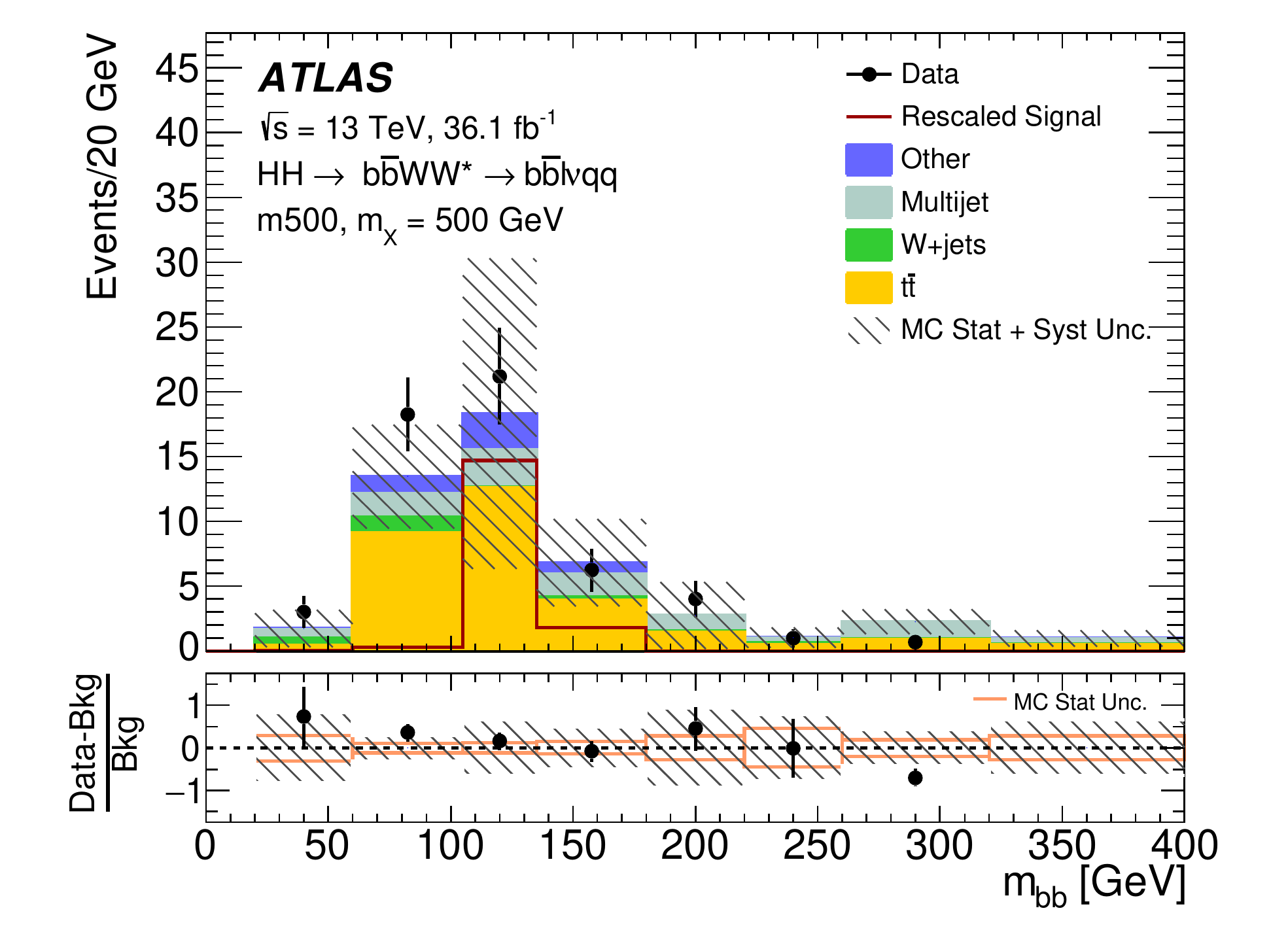}
\end{flushleft}
\caption{The $m_{b \bar{b}}$ distribution in the resolved analysis for the \emph{non-res} and \emph{m500} selections at the end of the
  selection sequence, before applying the $m_{b \bar{b}}$ requirement. The signals shown
  are from SM non-resonant $HH$ production scaled up by a factor of 300 (left) and from a scalar resonance with mass 500~\GeV\ scaled to the expected upper-limit
  cross section reported in Section~\ref{sec:MainResults} (right).  The lower panel shows the fractional difference 
   between data and the total expected background 
  with the corresponding statistical and total uncertainty. 
} \label{fig:mbb_1}
\end{figure}

\begin{figure}
\begin{flushleft}
\includegraphics*[width=0.45\textwidth]{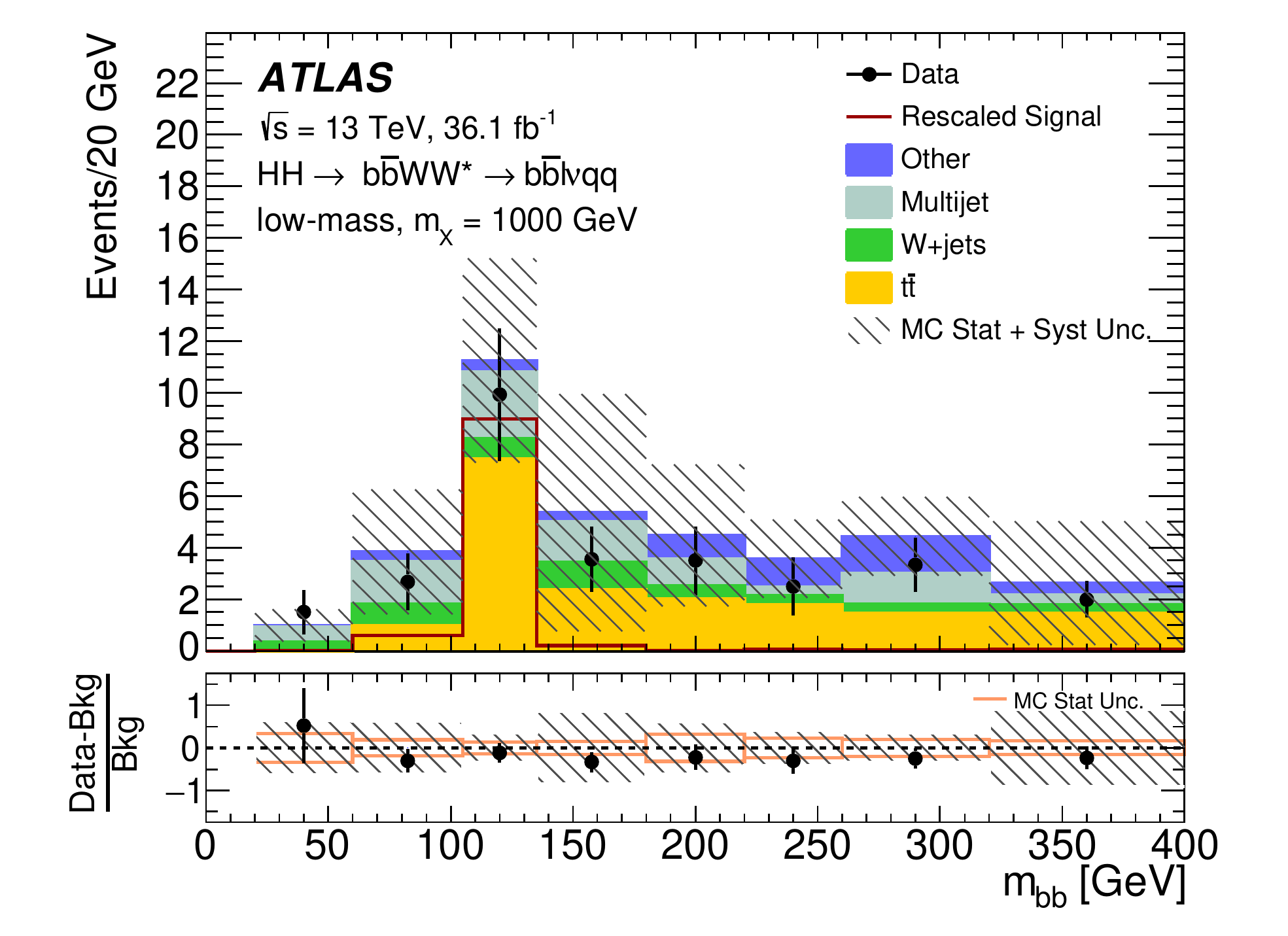}
\includegraphics*[width=0.45\textwidth]{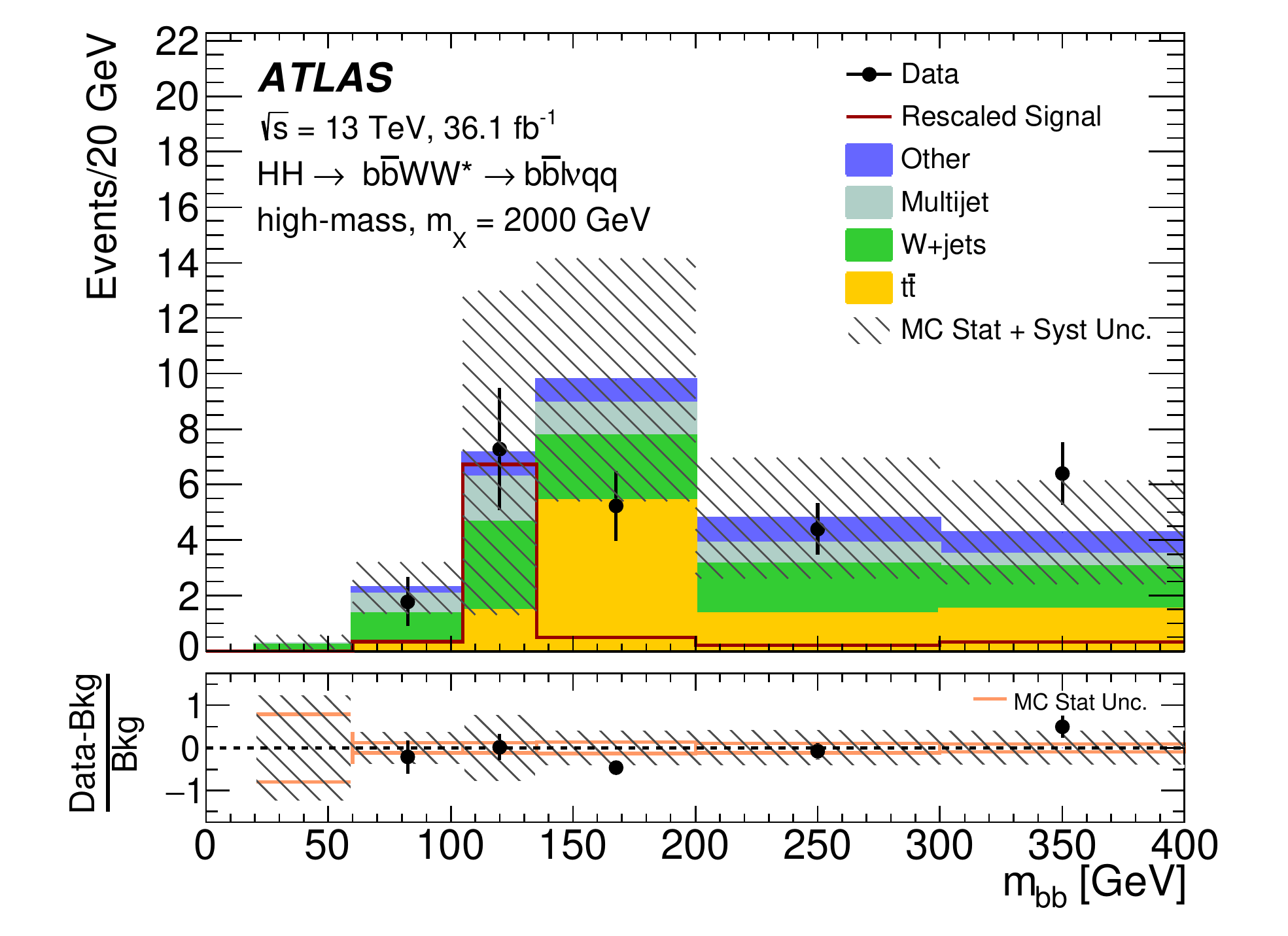}
\end{flushleft}
\caption{The $m_{b \bar{b}}$ distribution in the resolved analysis for the 
  \emph{low-mass} and \emph{high-mass} selections at the end of the
  selection sequence, before applying the $m_{b \bar{b}}$ requirement. The signals shown
  are from scalar resonances with mass 1000~\GeV\ (left) and 2000~\GeV\ (right)
  scaled to the expected upper-limit cross section reported in
  Section~\ref{sec:MainResults}.  The lower panel shows the fractional difference 
   between data and the total expected background 
  with the corresponding statistical and total uncertainty.} \label{fig:mbb_2}
\end{figure}

Figures~\ref{fig:mbb_1} and ~\ref{fig:mbb_2} show the $m_{b \bar{b}}$ distributions at the
selection stage where all requirements, including the $m_{HH}$ cut, are applied except the one on $m_{b \bar{b}}$ itself. 
The expected background is in agreement with the data over the entire distribution, and  
close to the signal region in particular. All simulated backgrounds are normalised according to their theoretical cross-sections, except \ttbar, which is normalised in the top CRs.

\subsection{Resolved analysis: systematic uncertainties}
\label{sec:systematics}

The main systematic uncertainties in the background estimate arise 
from the potential mis-modelling of background components.
For \ttbar background, MC simulation is used to derive the acceptances
in all analysis regions, while the normalisation is taken from the top control region and applied in the signal regions. Therefore, the acceptance ratio between signal and control regions is affected by theoretical uncertainties in the simulated \ttbar sample.
These  uncertainties are estimated by considering five sources: the
matrix element generator used for the $t \bar{t}$ simulation and the matching
scheme used to match the NLO matrix element with the parton shower, the
parton shower modelling, 
the initial-state (Initial State Radiation, ISR) and final-state  (Final State Radiation, FSR) gluon emission modelling,
the dependence on the choice of the PDF set and the dependence on the renormalisation and factorisation scales. Matrix element generator and matching systematic
uncertainties are computed by comparing samples generated by 
\textsc{\rm{a}MC@NLO}~\cite{Alwall2014} and \textsc{Powheg}, both interfaced with 
\textsc{Herwig++} for showering and fragmentation.
Parton shower systematic uncertainties 
are computed by comparing samples generated using \textsc{Powheg}+\textsc{Pythia6} 
and \textsc{Powheg}+\textsc{Herwig++}. 
Initial-state and final-state radiation systematic uncertainties are computed by varying the 
generator parameters from their nominal values to increase or decrease the amount of radiation.
The PDF uncertainties are computed using the eignevenctors of the CT10 PDF set. Uncertainties due to missing higher-order corrections, labelled scale uncertainties, are computed by independently scaling the renormalisation and factorisation scales in 
\textsc{\rm{a}MC@NLO}+\textsc{Herwig++} by a factor of two, while keeping
the renormalisation/factorisation scaling ratio between 1/2 and 2.
These systematic uncertainties are summarised in Table
\ref{tab:ttbarModeling}.

\begin{table}
\caption{Percentage uncertainties from \ttbar modelling on the \ttbar background contributions in all signal regions of the resolved analysis.}
\label{tab:ttbarModeling}
\centering
\begin{tabular}{l|c|c|c}
Source & \emph{non-res (\%)} &\emph{m500} and \emph{low-mass} (\%) &\emph{high-mass} (\%) \\\hline
Matrix element   &  7             &   0.5       &   4       \\
Parton shower    &  4             &  16      &   10       \\
ISR/FSR          & 15             &   5       &   8      \\
PDF              &  5             &   3	  &   6       \\
Scale            &  3             &   2       &   4       \\
\hline
Total              & 18             & 17            & 15\\
\end{tabular}
\end{table}

Uncertainties in the modelling of $W$+jets background are computed in
each signal region (SR) and top control region (CR). Three sources of 
uncertainty are considered: scale variation, PDF set variation
 and generator modelling uncertainties. Scale uncertainties 
are computed by scaling the nominal renormalisation 
and factorisation scales by a factor of two. PDF uncertainties are computed using 
the NNPDF \cite{Ball:2014uwa} error set,  
while generator modelling uncertainties are obtained 
by comparing the nominal \textsc{Sherpa}-generated
sample with a sample generated with \textsc{Alpgen}~\cite{Mangano:2002ea} and
showered with \textsc{Pythia6}~\cite{Sjostrand:2006za}. 
The values obtained in each region are summarised in Table \ref{tab:wplusjet_uncertainties}. 

\begin{table}
\caption{Theoretical percentage uncertainties on the predicted $W/Z$+jets event
  yield  in the top control regions and the signal regions for all selections.}
\centering
\begin{tabular}{l|cc|cc|cc}
Source            &\multicolumn{2}{c|} {\emph{non-res} (\%)} 	&
                                                                  \multicolumn{2}{c|}{\emph{m500
                                                                  and low-mass} (\%)}  		& \multicolumn{2}{c}{\emph{high-mass} (\%)}   \\
\hline
& SR & CR & SR & CR & SR & CR \\
\hline

Modelling/Parton Shower       & 40 & 40 	&   40 & 40 &  20 & 20             \\
PDF                  & 30 & 7 	&  40 & 10 & 30 & 20         \\
Scale                & 20 & 30 & 20 & 30 & 30 & 30      \\
\end{tabular}
\label{tab:wplusjet_uncertainties}
\end{table}

For the data-driven multijet background, three sources of uncertainty are identified. 
The non-closure correction term $F$ is computed using data at an early stage of the selection sequence, where 
contamination by the signal can be considered negligible. Its difference from 
the value obtained using a simulated multijet event sample is 40\%
and is assigned as an uncertainty in the multijet estimation. The $F$
value can be affected by the analysis selection requirements. A systematic
uncertainty (extrapolation uncertainty) is added by comparing the maximum variation among the $F$ values 
evaluated after each selection requirement. Finally, the uncertainty due to the dependence of the $F$ value on lepton flavour (flavour uncertainty) is computed
as the maximum difference between the nominal $F$ value and the $F$ value calculated for electrons and muons separately. 
The extrapolation (flavour) uncertainty is found to be 16\% (9\%) for 
the \emph{non-res} selection, 32\% (9\%) for the \emph{m500} and 
\emph{low-mass} resonant selections, and 45\% (6\%) for the \emph{high-mass} 
resonant selection.

Single-top-quark production is one of the smaller backgrounds in this analysis. Theoretical cross-section 
uncertainties vary from 5\% for associated $Wt$ production to
4\% for s- and t-channel single-top production. The largest of
these is conservatively assigned to all single-top production modes.  
Further modelling systematic uncertainties are calculated 
by employing the difference between the nominal sample using the Diagram Removal scheme described in Ref.~\cite{Frixione:2008yi} and a 
 sample using the Diagram Subtraction scheme for the 
dominant single-top production mode, $Wt$. The uncertainties are  50\%, 
 for the \emph{non-res}, \emph{m500} and \emph{low-mass} analyses, and 80\%
for the \emph{high-mass} analysis. 

Systematic uncertainties in the signal acceptance are computed 
by varying the renormalisation and factorisation
scales with a variation of up to a factor of two, and using the same procedure as
for the $t\bar{t}$ background.
PDF uncertainties are computed using PDF4LHC15\_30~\cite{Butterworth:2015oua} PDF sets, which 
include the envelope of three PDF sets, namely CT14, MMHT14,
NNPDF3.0. The resulting uncertainties are less than 1.1\% for the scale
and less than 1.3\% for the PDFs. Parton shower uncertainties are computed by 
comparing the \textsc{Herwig++} showering with that of
\textsc{Pythia8}, and this results in less than 2\% uncertainty.

The detector-related systematic uncertainties affect both the
background estimate and the signal yield. In this analysis the largest of these
uncertainties are related to the jet energy scale (JES), jet energy resolution
(JER), $b$-tagging efficiencies and mis-tagging rates. The JES
uncertainties for the small-$R$ jets are derived from $\sqrt{s} = 13~\TeV$ 
data and simulations \cite{Aaboud:2017jcu}, while the JER uncertainties are extrapolated
from 8~\TeV\ data using MC simulations \cite{ATL-PHYS-PUB-2015-015}. 
The uncertainty due to $b$-tagging is evaluated following the procedure described in Ref.~\cite{Aaboud:2018xwy}. 
The uncertainties associated with lepton reconstruction and energy
measurements have a negligible impact on the final results. All lepton
and jet measurement uncertainties are propagated to the calculation of
$\met$, and additional uncertainties are included in the
scale and resolution of the soft term. The overall impact of the
$\met$ soft-term uncertainties is also small. Finally, the uncertainty in the combined integrated luminosity is $3.2\%$ ~\cite{DAPR-2013-01}.

\section{Boosted analysis}
\subsection{Boosted analysis: event selection}
\label{sec:boosted_selection}

As in the resolved analysis, data used in the boosted analysis were recorded by single-lepton triggers, and only  events that contain at least one reconstructed electron or muon matching the trigger lepton candidate are analysed. Requirements on \pt, $|d_0|/\sigma_{d_0}$ and $|z_0\sin\theta|$ of the lepton tracks are also the same as in the resolved analysis. 

Events are required to have at least one large-$R$ jet with an angular distance $\Delta R$ > 1.0 from the reconstructed lepton. The highest-\pt large-$R$ jet is identified as the  $H\to b\bar{b}$ candidate. The large-$R$ jet mass is required to be between 30~\GeV\ and 300~\GeV. In order to reconstruct the $H \to WW^{\ast}$ system, events with at least two small-$R$ jets 
with an angular  distance  $\Delta R$ > 1.4 from the $H \to b\bar{b}$ candidate are selected. 
The hadronically and leptonically decaying $W$ bosons are then reconstructed following the same algorithm as in the resolved analysis. In order to reduce the $t\bar{t}$ background, events are rejected if they contain any small-$R$ jet passing the $b$-tagging requirement.

Signal regions (SR) are defined with at least two associated
track-jets within the large-$R$ jet and requiring that the two highest-\pt track-jets are also $b$-tagged. The
large-$R$ jet mass must be between 90~\GeV\ and 140~\GeV. An additional requirement of $\met$ > 50~\GeV\ is imposed to reject multijet backgrounds. 

In order to assess the modelling of the dominant \ttbar background, a validation region (VR) is defined outside the large-$R$ jet 
signal region mass window and labelled top VR. Any event with a large-$R$ jet mass $m_\text{Large-$R$ jet} < 90~\GeV$ or $m_\text{Large-$R$ jet} > 140~\GeV$ falls in the top VR. By construction, the top VR is orthogonal to the SR.

\subsection{Boosted analysis: background determination}
\label{sec:boosted_backgrounds}
In the boosted analysis the presence of a signal is indicated by an excess of events 
above the SM prediction of the  background  $m_{HH}$
distribution at the end of the event selection. 
Similarly to the resolved analysis, the $t\bar{t}$ process is the dominant background. Therefore, 
a dedicated validation region is used to check its modelling as
defined in Section~\ref{sec:boosted_selection}. 
The event yields from \ttbar, $W/Z$+jets, single-top-quark and diboson processes in the 
signal region and the top VR are modelled using
simulation and normalised to the expected SM cross section described in Section~\ref{sec:samples}.

The multijet component of the background
is estimated using the  data-driven method as in the resolved
analysis. In the boosted analysis a higher requirement on $\met$ ($\met > 50~\GeV$) is applied,  while the
 cut on $|d_0|/\sigma_{d_0}$  is the
 same. For the boosted analysis, the correlation between
$|d_0|/\sigma_{d_0}$ and \met is estimated in multiple MC
 background samples and also in data, and it is found to be
 negligible. Hence, the multijet yield in region A can be
estimated using the relation $N_{\text{A}} = N_{\text{C}} N_{\text{B}}/N_{\text{D}}$.  The multijet estimation  is performed separately for
 the muon and the electron channel. 
The $N_{\text{B}}/N_{\text{D}}$ ratio is calculated inclusively in the large-$R$ jet mass distribution.
The $m_{HH}$ distribution of the multijet background is estimated by subtracting the prompt-lepton MC backgrounds from the data in the 1-tag region, where the 1-tag region is defined as the region where all selections are applied except that the large-$R$ jet is required to have only one track-jet tagged as a $b$-jet. 

\begin{table}
\caption{Predicted and observed event yields in the top VR for the boosted analysis. The uncertainty shown for the multijet background is due to the number of data events
in the C region. For all other backgrounds the uncertainties are due to the 
finite MC sample sizes.}
\begin{center}
\begin{tabular}{l|r@{}c@{}l}
Process       &  \multicolumn{3}{c}{Events}  \\
\hline 
\ttbar    &  1000&$\,\pm\,$&21   \\
$W$+jets        &  570&$\,\pm\,$&10    \\
Multijet           & 380&$\,\pm\,$&20    \\ 
Single top    &  160&$\,\pm\,$&7     \\ 
Dibosons      &  40&$\,\pm\,$&3     \\
$Z$+jets        &  56&$\,\pm\,$&2     \\  
Background sum    &  2206&$\,\pm\,$\,&31    \\
\hline
Data          &  2179 & &   \\
\end{tabular} 
\end{center}
\label{tab:boosted_bkgd_mbbcr_yields}
\end{table}

\begin{figure}
\includegraphics*[width=0.48\textwidth]{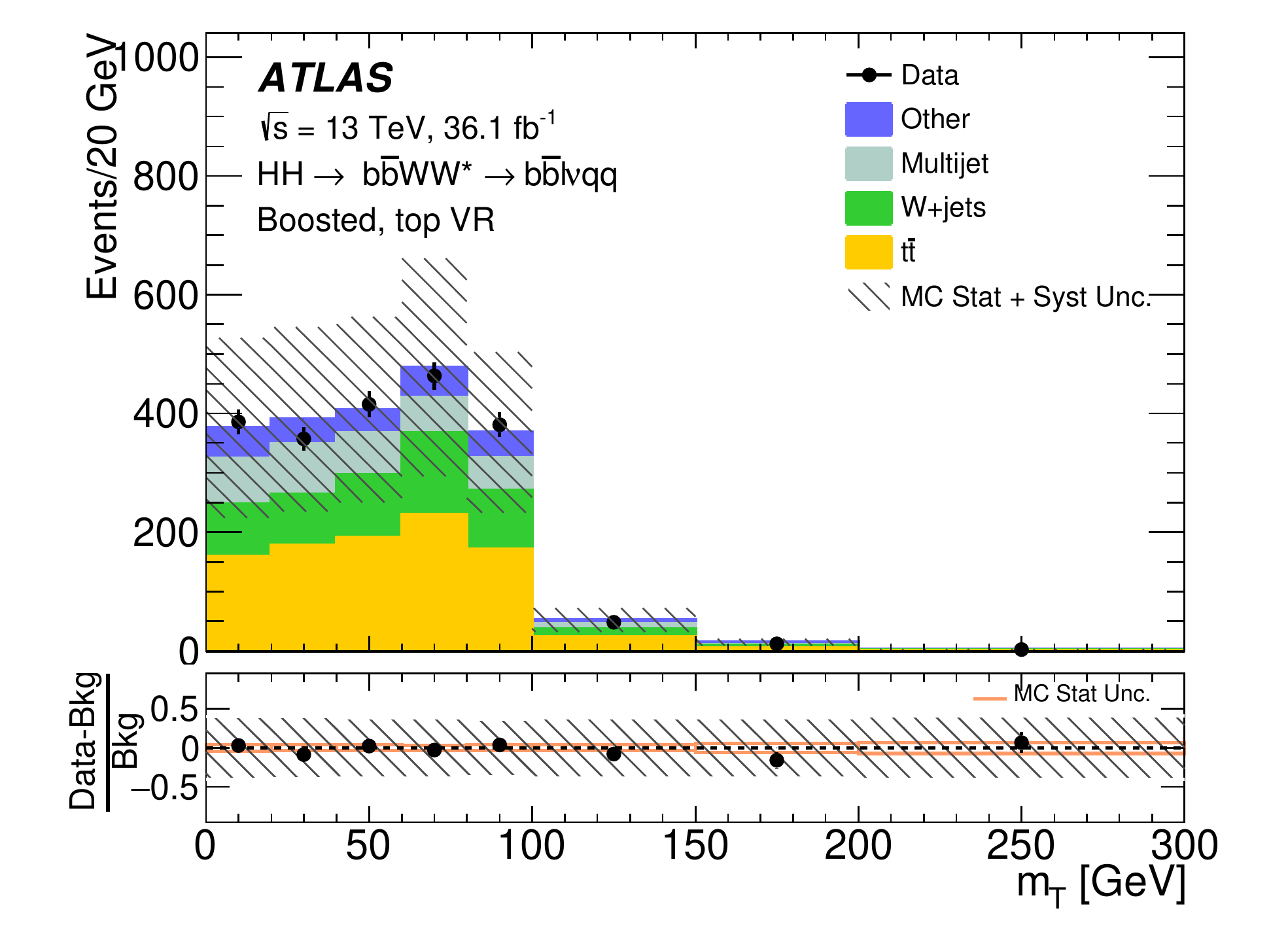}
\includegraphics*[width=0.48\textwidth]{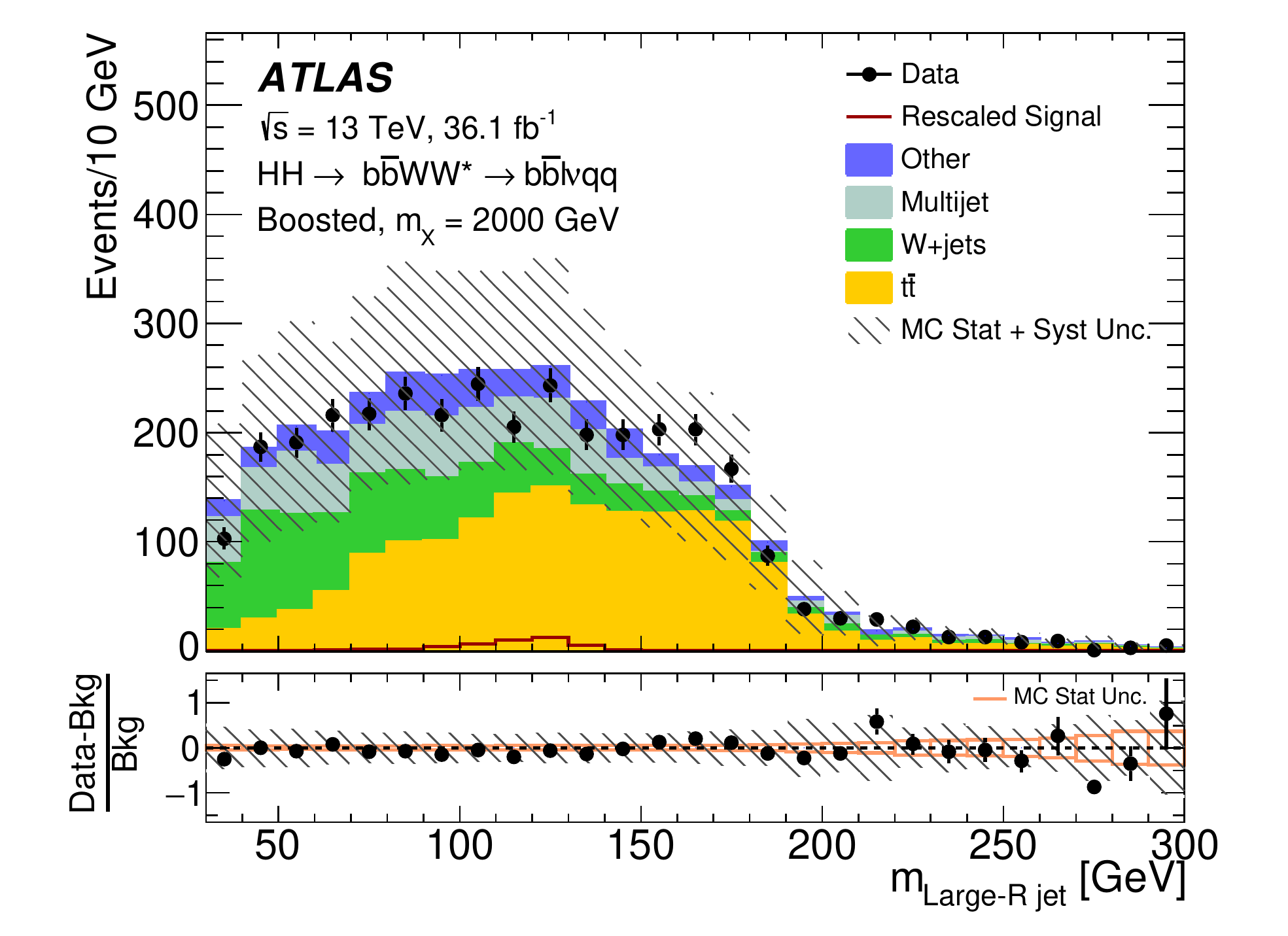}
\caption{The $\mT$ distribution (left) in the top VR, and inclusive $m_{\text{Large-$R$ jet}}$ distribution (right)
  after applying all selections.  
The signal distribution is negligible in the left plot, while in the right plot it has been scaled to the expected upper-limit 
cross section reported in Section~\ref{sec:MainResults}. The lower panel shows the fractional difference between data  
and the total expected background with the corresponding statistical and total
uncertainty.}
  \label{fig:boosted_mtlep_mbb}
\end{figure}

The modelling of the background is checked in the top VR.
Table~\ref{tab:boosted_bkgd_mbbcr_yields} reports the numbers of observed
and predicted background events in the top VR, showing good
agreement between the two. In order to check the validity of the
multijet background determination, 
the $\mT$ distribution is shown in Figure~\ref{fig:boosted_mtlep_mbb}. This variable is particularly sensitive to the multijet background
contamination. Additionally, the  $m_{\text{Large-$R$ jet}}$ variable used to define the
signal region and the top VR is shown in the same figure. The data and predicted background agree well, which builds 
confidence in the estimated efficiency of the $m_{\text{Large-$R$ jet}}$ requirement for signal and background.

\subsection{Boosted analysis: systematic uncertainties}
\label{sec:boosted_systematics}

The evaluation of detector modelling uncertainties in the boosted analysis follows the same approach as in the resolved analysis. The significant additions to those described in Section \ref{sec:systematics} are the uncertainties related to the large-$R$ jets. The large-$R$ jet energy resolution and scale, and jet mass resolution and scale uncertainties are derived $\it {in~situ}$ from 8~\TeV\  $pp$ collision data, taking into account MC simulation extrapolations for the different detector and beam conditions present in 8 and 13~\TeV\ data-taking periods~\cite{ATL-PHYS-PUB-2015-033}. The uncertainty in the $b$-tagging efficiency for track-jets is evaluated with the same method used for resolved calorimeter jets. The impact of these uncertainties on the final fit are shown in Table~\ref{tab:boosed_nplist_systematics}.

All SM backgrounds, except multijet, are modelled using MC simulation.  Therefore, predicted yields in both the signal and the top validation regions are affected by theoretical uncertainties. These uncertainties are computed following the same procedure as in the resolved analysis for $\ttbar$, $W/Z$+jets, single-top-quark and diboson backgrounds. For the $t\bar{t}$ background in the signal region, the uncertainties are summarised in Table~\ref{tab:boosted_ttbarModeling}. The uncertainties on single top quark production range from 20\% for ISR/FSR to 70\% stemming from the difference between the diagram removal and diagram subtraction schemes. Uncertainties in the modelling of $W/Z$+jets background range from 10\% stemming from PDF uncertainties to 45\% stemming from scale uncertainties. Diboson processes have a negligible impact on the total background.

\begin{table}[htbp!]
\caption{Uncertainties from different sources in the predicted yield of the $t\bar{t}$ background in the signal region of the boosted analysis.} 
\begin{center}
\begin{tabular}{l|c}
Source  &  Uncertainty (\%) \\ 
\hline
Matrix element    	&      		7.1 \\
Parton shower   	&                 7.8 \\
ISR/FSR      		&                 8.4 \\
PDF	        	  		 &                 1.9 \\
Scale	     		&                 5.0 \\
\hline
Total      			&                14.5~~ \\
\end{tabular}
\end{center}
\label{tab:boosted_ttbarModeling}
\end{table}

For the normalisation of the multijet background predicted in region A (See Section~\ref{sec:boosted_backgrounds}), several sources of uncertainty are considered. 
The uncertainties in the normalisation of $t\bar{t}$ and $W/Z$+jets in regions B, C and D contribute a systematic uncertainty of 25\% and 30\% respectively.
The relative difference between the large-$R$ jet mass acceptance in the 1-tag region C and in the 2-tag region C accounts for 15\%.
The propagation of the statistical uncertainty in the multijet yield in region C and the uncertainty in the 
$N_{\text{B}}/N_{\text{D}}$ ratio contribute about 23\%. 
The propagation of detector modelling systematic uncertainties, including the modelling uncertainty of the 
$|d_0|/\sigma_{d_0}$ requirement and of the MC backgrounds with prompt leptons subtracted from data in 
regions B, D and C, contribute about 45\%.
As an additional check on the prediction of the multijet yield with the ABCD method, a conditional
background-only likelihood fit of the large-$R$ jet mass distribution is performed in the VR. The difference between the multijet yield estimated with this method and the ABCD prediction is assigned as an uncertainty. This error accounts for  ~23\% of the total uncertainty in the multijet estimation.
All different sources of uncertainty are treated as independent and added  in quadrature for the final uncertainty of 80\% in the multijet normalisation.

For the simulated backgrounds, the systematic uncertainty in the $m_{HH}$ distribution shape is determined by comparing the nominal MC sample with the corresponding alternative (variation) MC samples described in Section~\ref{sec:systematics}.
The shape systematic uncertainty is determined by fitting a first-order polynomial to the ratio of the variation $m_{HH}$ distribution to the nominal $m_{HH}$ distribution, while keeping the same normalisation.
For the data-driven multijet background, the uncertainty in the $m_{HH}$ distribution shape is determined by 
comparing the shapes in the 2-tag and 1-tag C regions. 

Systematic uncertainties in the signal acceptance are computed following the same algorithm as the resolved analysis. The resulting uncertainties are less than 0.5\% for uncertainties due to missing higher-order corrections (labelled scale), less than 0.5\% for those due to PDFs, and approximately 2\% (5\%) in the lower (higher) mass range for those due to the parton shower.

\section {Results}
\label{sec:MainResults}
\subsection{Resolved analysis}
\label{sec:results}
\begin{figure}
\begin{center}
\includegraphics[width=0.45\textwidth]{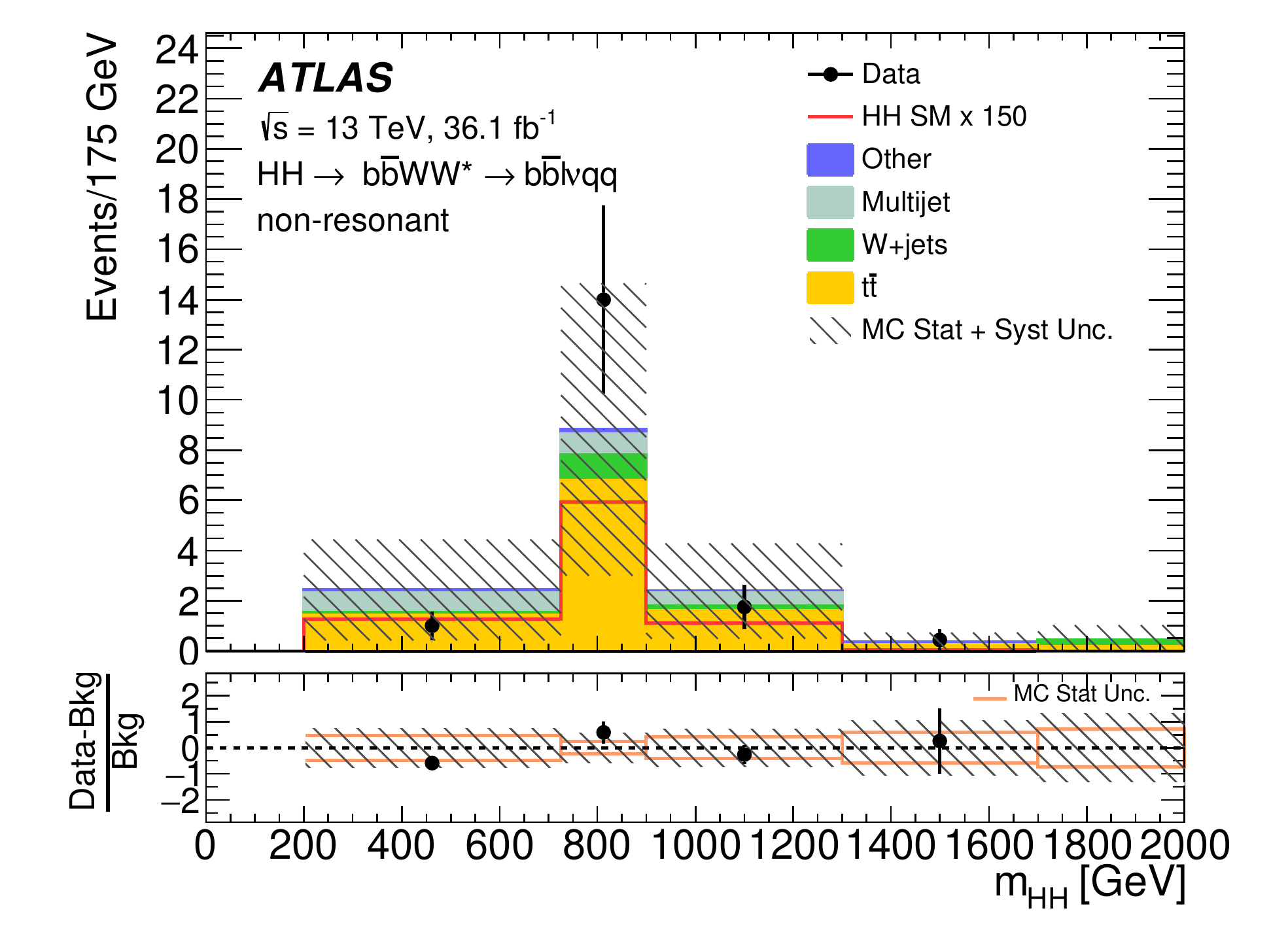} 
\includegraphics[width=0.45\textwidth]{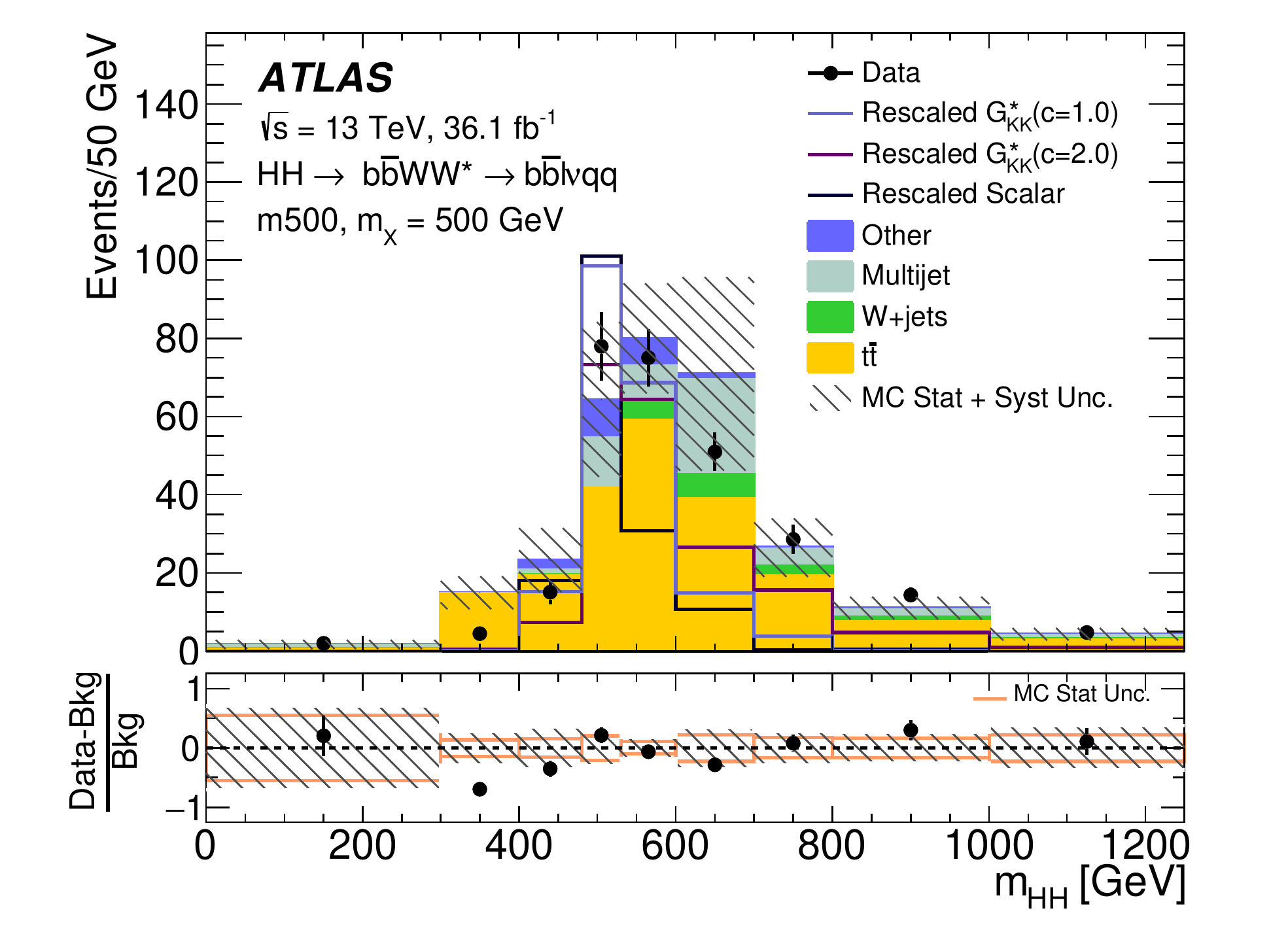} 
\end{center}
\caption{$m_{HH}$ distributions for non-resonant and \emph{m500} selections in the resolved analysis. For each selection the corresponding
  signal hypothesis, non-resonant, scalar resonance, and graviton with
  $c=1.0$ and $c=2.0$, is shown. For scalar and graviton signals, resonances with mass 500~\GeV\ are shown. The lower panel shows the fractional difference between data and the total expected background 
  with the corresponding statistical and total
  uncertainty. The non-resonant signal is multiplied by a factor of 150 with respect to 
  the expected SM cross section.  The scalar signal is multiplied 
  by a factor of 5,  the graviton $c=1.0$ by a factor of 5 and the graviton
  $c=2.0$ by a factor of 1 with respect to the expected upper-limit cross section
  reported in Section~\ref{sec:MainResults}. }
\label{fig:mhh_1}
\end{figure}

\begin{figure}
\begin{center}
\includegraphics[width=0.45\textwidth]{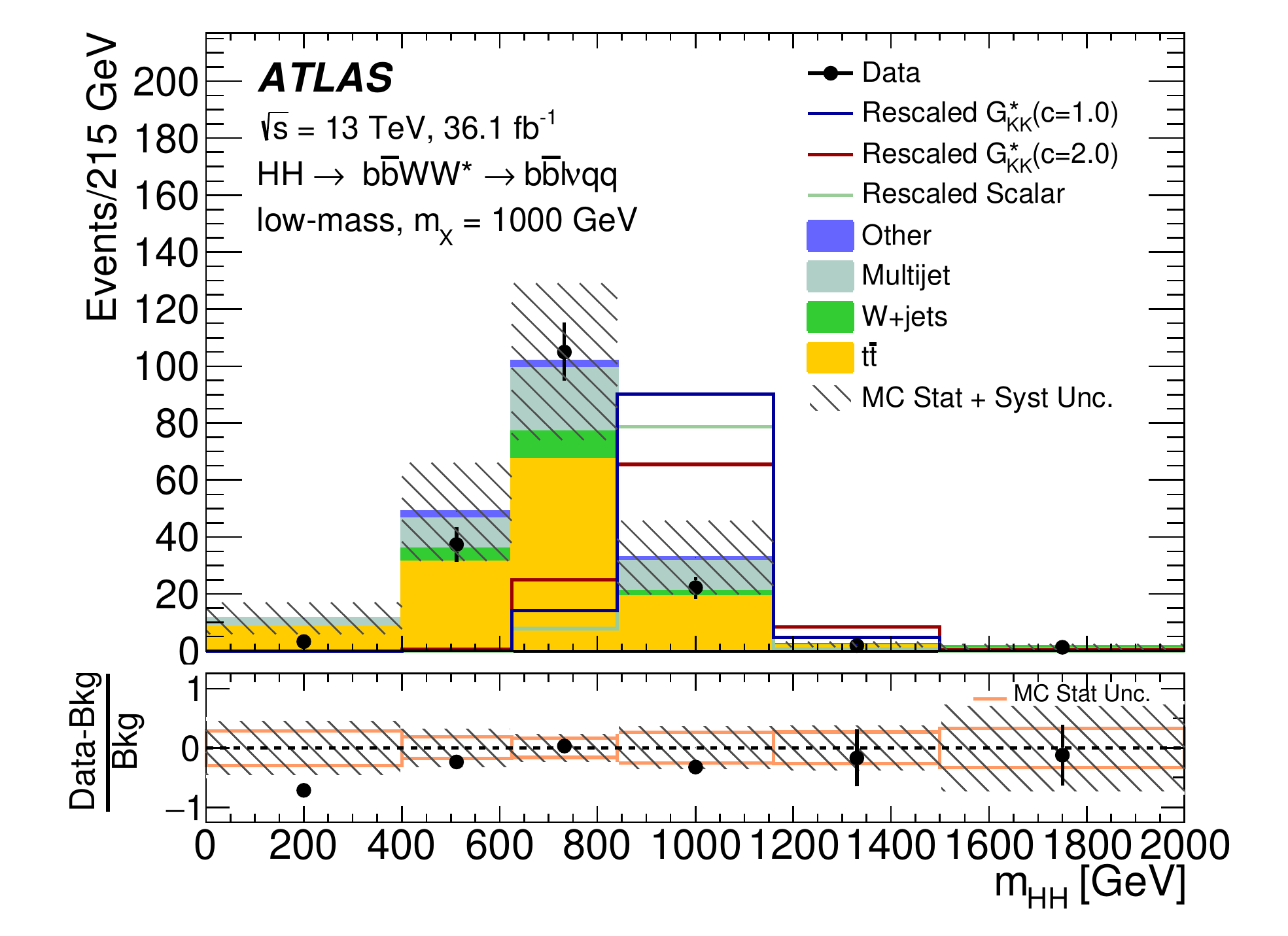} 
\includegraphics[width=0.45\textwidth]{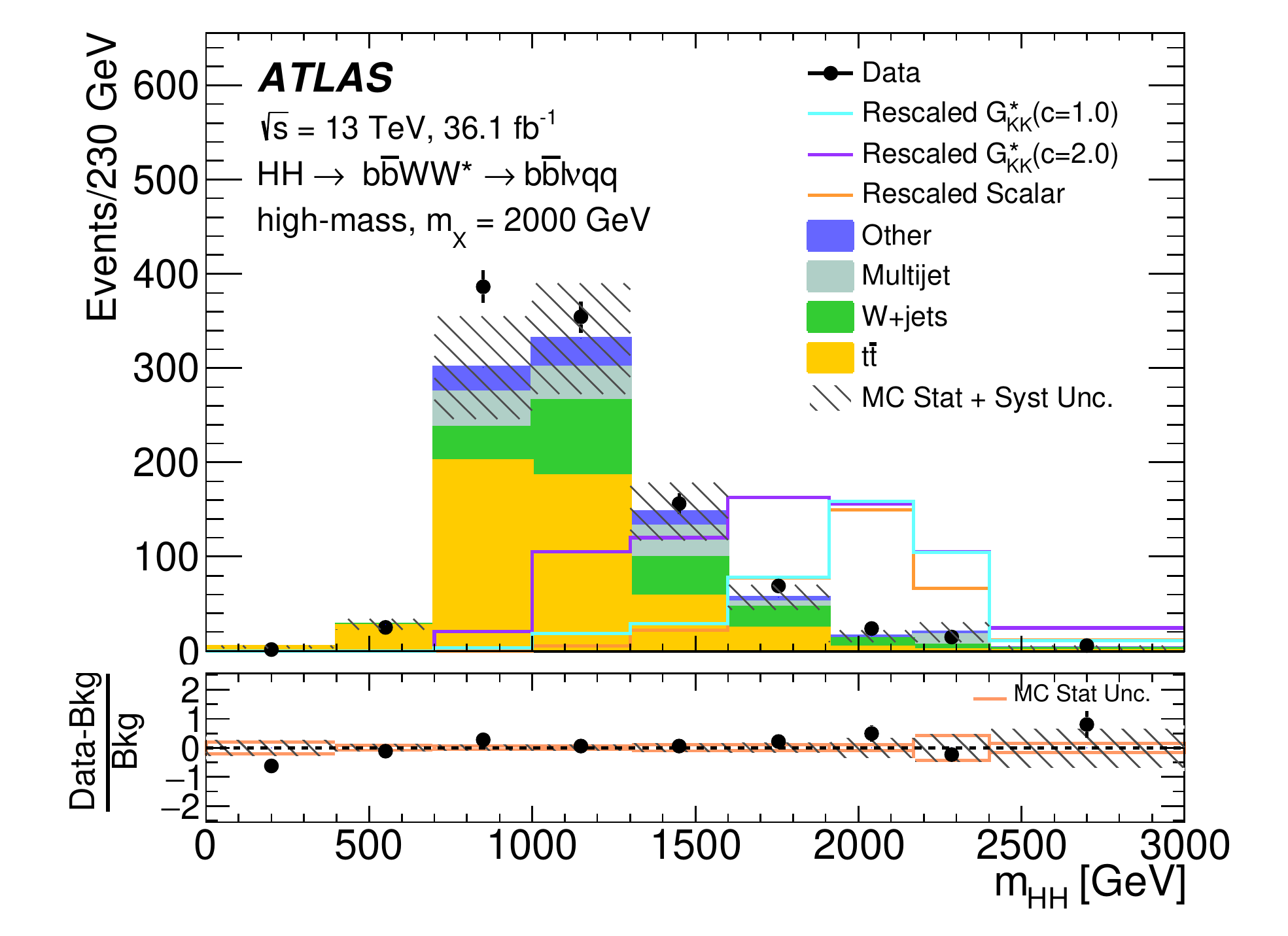} 
\end{center}
\caption{$m_{HH}$ distributions in the resolved analysis selections. For each selection the corresponding
  signal hypothesis, scalar resonance, and graviton with
  $c=1.0$ and $c=2.0$,  and mass 1000 (2000)~\GeV\ for the  \emph{low-mass} (\emph{high-mass}) analysis, are shown.  The lower panel shows the fractional difference 
   between data and the total expected background 
  with the corresponding statistical and total
  uncertainty. In the plot on the left the scalar signal is multiplied by a factor of 8, the
  graviton $c=1.0$ by a factor of 10 and the graviton $c=2.0$ by a
  factor of 2 with respect to the expected upper-limit cross section
  reported in Section~\ref{sec:MainResults}; for the plot on the right
the multiplying factors are 20 for the scalar signal, 10 for  the graviton
$c=1.0$  signal and 5 for the graviton $c=2.0$ signal.}
\label{fig:mhh_2}
\end{figure}

After applying the requirements listed in Table \ref{tab:sig_reg_summary},  the invariant mass of the $HH$ system ($m_{HH}$) is distributed as shown in Figures \ref{fig:mhh_1} and \ref{fig:mhh_2}.
Data are generally in good agreement with the expected
background predictions within the total uncertainty.
The signal $m_{HH}$ distribution is shown in the figure for the non-resonant, the scalar resonance,
and the two graviton hypotheses with $c = 1.0$ and $c=2.0$. 
The scalar samples are simulated in the narrow-width approximation, so the 
reconstructed width is exclusively due to the detector
resolution. The same holds for graviton samples with $c=1.0$, while
$c=2.0$ graviton samples have a significant intrinsic width that
leads to a loss of sensitivity.

The $m_{HH}$ distribution is sampled with resonance-mass-dependent
$m_{HH}$ requirements as reported in Table \ref{tab:mhh_cuts}.
The numbers of events in the signal and control regions (the $t\bar{t}$
control region and the C region of the multijet estimation procedure) are
simultaneously fit using a  maximum-likelihood approach. The fit includes six contributions: signal, $W$+jets, 
$Z$+jets, $t\bar{t}$, single-top-quark production, diboson and multijet. The
$t\bar{t}$ and multijet normalisations are free to float, the C
region of the ABCD method being directly used in the fit, while the
diboson, $W$+jets and $Z$+jets backgrounds are constrained
to the expected SM cross sections within their uncertainties. 

The fit is performed after combining the electron and muon channels. Statistical uncertainties due to the limited
sample sizes of the simulated background processes are taken into
account in the fit by means of nuisance parameters, which are
parameterised by Poisson priors. Systematic uncertainties are taken into account as
nuisance parameters with Gaussian constraints. For each
source of systematic uncertainty, the correlations across bins and
between different kinematic regions, as well as those between signal
and background, are taken into account. 
Table~\ref{tab:event_yields_low} shows
the post-fit number of predicted backgrounds, observed data, and the
signal events normalised to the expected upper limit cross
sections.

\begin{table}
\caption{Data event yields, and post-fit signal and background event yields in the final signal region for the non-resonant analysis and the resonant analysis in the 500--3000~\GeV\ mass range. The errors shown
 are the MC statistical and systematic uncertainties described in
 Section~\ref{sec:systematics}. The yields are shown for three signal
 models: a scalar ($S$) and two Randall--Sundrum gravitons with $c=1.0$
 and $c=2.0$ ($G^{\ast}_{\rm KK}$). Signal event yields are normalised to the expected upper-limit cross section.}  
\label{tab:event_yields_low}
\small
\begin{center}
\begin{tabular}{r|*{4}{r@{}c@{}l|}r}
\hline
\multicolumn{14}{c}{Resonant analysis} \\
\hline
$m_X$ [GeV] & \multicolumn{3}{c}{$S$} & \multicolumn{3}{c}{$G^{\ast}_{\rm KK}$ $(c=1.0)$} & \multicolumn{3}{c}{$G^{\ast}_{\rm KK}$ $(c=2.0)$} & \multicolumn{3}{c}{Total Bkg.} & Data \\
\hline
500~~~~~~& 18&$\,\pm\,$&5 & 20&$\,\pm\,$&5 & 18&$\,\pm\,$&5 & 19&$\,\pm\,$&6 & 26 \\
600~~~~~~& 13&$\,\pm\,$&2 & 15&$\,\pm\,$&2 & 13&$\,\pm\,$&2 & 17&$\,\pm\,$&6 & 16 \\
700~~~~~~& 16&$\,\pm\,$&2 & 17&$\,\pm\,$&2 & 16&$\,\pm\,$&2 & 25&$\,\pm\,$&8 & 22 \\
750~~~~~~& 20&$\,\pm\,$&2 & 22&$\,\pm\,$&2 & 20&$\,\pm\,$&2 & 22&$\,\pm\,$&9 & 27 \\
800~~~~~~& 18.4&$\,\pm\,$&1.5 & 19.7&$\,\pm\,$&1.6 & 18.2&$\,\pm\,$&1.5 & 20&$\,\pm\,$&8 & 28  \\
900~~~~~~& 16.3&$\,\pm\,$&1.6 & 17.0&$\,\pm\,$&1.7 & 16.1&$\,\pm\,$&1.6 & 20&$\,\pm\,$&7 &
                                                                      23
  \\
1000~~~~~~& 12.0&$\,\pm\,$&1.3 & 12.3&$\,\pm\,$&1.4 & 11.9&$\,\pm\,$&1.3 & 14&$\,\pm\,$&5 &
                                                                       11
  \\
1100~~~~~~& ~~9.6&$\,\pm\,$&1.2 & ~~9.8&$\,\pm\,$&1.2 & ~~9.5&$\,\pm\,$&1.1 & ~~8&$\,\pm\,$&3 & 8
  \\
1200~~~~~~& ~~8.1&$\,\pm\,$&0.9 & ~~8.2&$\,\pm\,$&0.9 & ~~8.1&$\,\pm\,$&0.9 & ~~6&$\,\pm\,$&3 & 5
  \\
1300~~~~~~& ~~5.1&$\,\pm\,$&0.7 & ~~5.1&$\,\pm\,$&0.7 & ~~6.2&$\,\pm\,$&0.8 & ~~3.5&$\,\pm\,$&1.8 &
                                                                       1
  \\
1400~~~~~~& 4.3&$\,\pm\,$&0.3 & 4.1&$\,\pm\,$&0.3 & 4.0&$\,\pm\,$&0.3 & ~~1.1&$\,\pm\,$&0.2 & 0 \\
1500~~~~~~& 3.5&$\,\pm\,$&0.3 & 3.5&$\,\pm\,$&0.3 & 3.5&$\,\pm\,$&0.3 & 1.1&$\,\pm\,$&0.2 & 0 \\
1600~~~~~~& 3.1&$\,\pm\,$&0.3 & 3.1&$\,\pm\,$&0.3 & 3.2&$\,\pm\,$&0.3 & ~~0.4&$\,\pm\,$&0.3 & 1 \\
1800~~~~~~& 14.1&$\,\pm\,$&1.8 & 14&$\,\pm\,$&2 & 14&$\,\pm\,$&2 & 17&$\,\pm\,$&5 & 21 \\
2000~~~~~~& 8.7&$\,\pm\,$&1.0 & 8.9&$\,\pm\,$&1.0 & 8.8&$\,\pm\,$&1.0 & 8&$\,\pm\,$&3 & 9
  \\
2250~~~~~~& 7.9&$\,\pm\,$&1.1 & 8.2&$\,\pm\,$&1.2 & 8.2&$\,\pm\,$&1.2 & 6&$\,\pm\,$&2 & 7
  \\
2500~~~~~~& 5.5&$\,\pm\,$&0.8 & 5.6&$\,\pm\,$&0.8 & 5.6&$\,\pm\,$&0.8 & 3.3&$\,\pm\,$&1.4 &
                                                                       3\\
2750~~~~~~& 5.7&$\,\pm\,$&1.0 & 6.1&$\,\pm\,$&1.1 & 6.0&$\,\pm\,$&1.1 & 3.1&$\,\pm\,$&1.3 &
                                                                       3\\
3000~~~~~~& 4.3&$\,\pm\,$&0.7 & 4.6&$\,\pm\,$&0.7 & 4.5&$\,\pm\,$&0.7 & 2.1&$\,\pm\,$&1.0 &
                                                                       1 \\
\hline
\multicolumn{14}{c}{Non-resonant analysis} \\
\hline
\multicolumn{10}{c|}{Rescaled SM signal} & \multicolumn{3}{c}{Total Bkg.} & Data \\
\hline
\multicolumn{10}{c|}{17$\,\pm\,$2} & 21&$\,\pm\,$&8 & 22\\
\end{tabular}
\end{center}
\end{table}

No significant excess over the expectation is observed and the results are used to evaluate an upper limit at the $95\%$
confidence level (CL)  on the production cross section times the
branching fraction for the signal hypotheses under consideration.
 The exclusion limits are calculated with a modified frequentist
method~\cite{CLs_2002}, also known as CL$_{\mathrm{s}}$, and the
profile-likelihood test statistic~\cite{asymptotics}.  None of the considered
systematic uncertainties is significantly constrained or
pulled in the likelihood fit. 

In the non-resonant signal hypothesis the observed (expected) upper
limit on the $\sigma(pp \to HH) \times {\mathcal{B}}( HH \to bbWW^{\ast})$ at 95\%
CL is:
\[
\sigma(pp \to HH) \cdot {\mathcal{B}}(HH \to bbWW^{\ast}) < 2.5 \, \left
  (2.5^{+1.0}_{-0.7} \right )  \,
{\mathrm{pb}}.
\]
The branching fraction ${\mathcal{B}}(HH \to bbWW^{\ast}) = 2\times
{\mathcal{B}}(H \to bb) \times {\mathcal{B}}(H \to WW^{\ast}) = 0.248$ is used to 
obtain the following observed (expected) limit on the $HH$ production cross section at 95\% CL: 
\[
\sigma(pp \to HH) < 10 \, \left (10^{+4}_{-3} \right) \,  {\text{pb,}}
\]

which corresponds to 300~(300$^{+100}_{-80}$) times the SM predicted cross section.
Including only the statistical uncertainty, the expected upper limit for the non-resonant production is
190 times the SM prediction.
This result, when compared with other $HH$ decay channels, is not competitive.  This is mainly  due to the similarity of the reconstructed $m_{HH}$ spectrum between the non-resonant SM signal
and the $t\bar{t}$ background that makes the separation between the
two processes difficult. 

\begin{figure}[!h]
\begin{center}
\includegraphics*[width=0.45\textwidth]{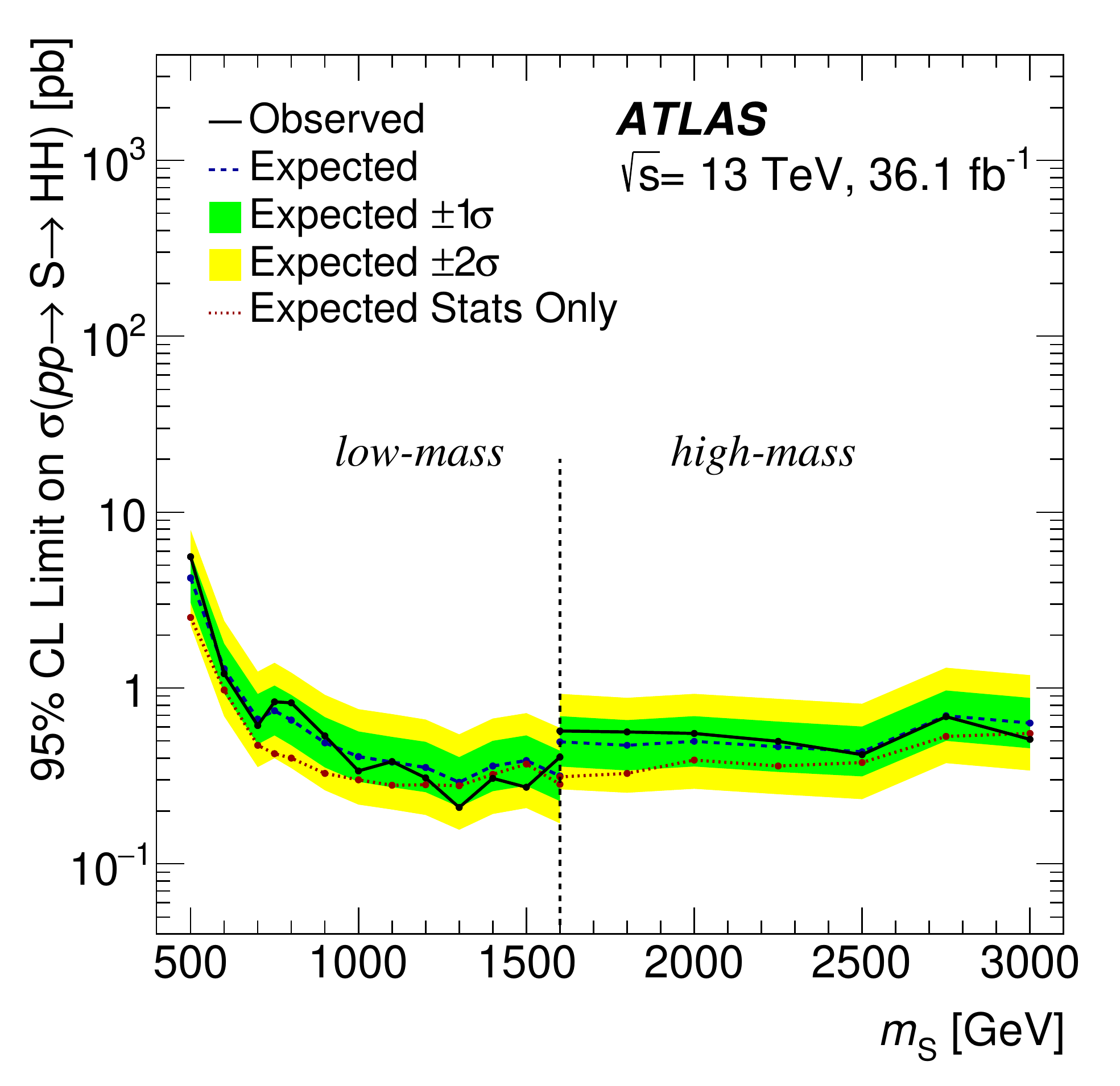}
\includegraphics*[width=0.45\textwidth]{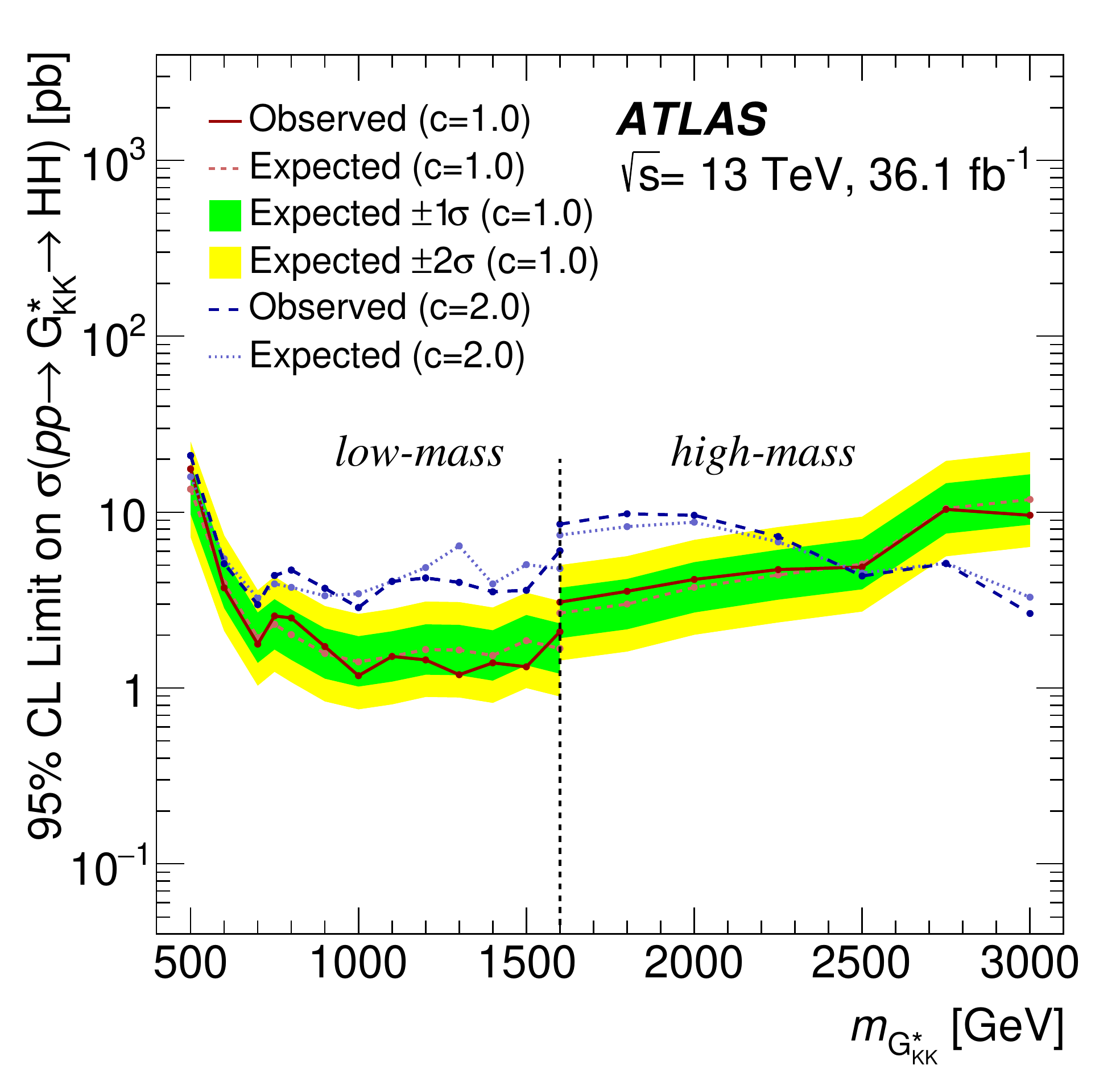}
\caption{Expected and observed upper limit at 95\% CL on the cross section of resonant pair
         production for the resolved analysis in the heavy scalar boson $S$ model (left) and the
         spin-2 graviton model in two $c$ parameter
         hypotheses (right). The left plot also shows the expected limit
         without including the systematic errors in order to show
         their impact. The impact of systematic errors is similar for the graviton models. 
          }
\label{fig:limit}
\end{center}
\end{figure}

Figure~\ref{fig:limit} shows the expected and observed limit
curves for the production cross section of a scalar $S$ and graviton $G^{\ast}_{\rm KK}$ particle. 
The graviton case is studied for the two values of the model parameter
$c$ described previously.

The analysis is most sensitive for a mass value of 1300~\GeV\ with an
expected upper limit of 0.35~pb on $\sigma(pp \to HH)$. At this mass the observed exclusion limit  is 0.2~pb. 
In both the non-resonant and resonant cases, the impact of the
systematic uncertainties is observed to be large.
In order to quantify the impact of the
systematic uncertainties, a fit is performed where the estimated
signal yield,  normalised to an arbitrary cross-section value, is multiplied
by a scaling factor $\alpha_{\mathrm{sig}}$, which is treated as
the parameter of interest in the fit.
The fit is performed using pseudo-data and the contribution to the  uncertainty in
$\alpha_{\mathrm{sig}}$ from several sources is determined. The contribution of the statistical uncertainty  to the total
uncertainty in $\alpha_{\mathrm{sig}}$, shown in Table~\ref{tab:pre-fit-stat},  is decomposed into
signal region statistics, top CR statistics and multijet CR
statistics. The contribution of the systematic uncertainties to the
total uncertainty is decomposed into the dominant components and shown in Table~\ref{tab:pre-fit-systematics}. The dominant
systematic uncertainties vary across the mass range, but some of the most relevant ones are due to $t \bar{t}$ modelling, $b$-tagging systematic uncertainties, and those related to jet measurements.

\begin{table}
\caption{Statistical contribution (in percentage) to the total error
  in the scaling factor $\alpha_{\mathrm{sig}}$ for
  the non-resonant signal and three scalar-signal mass hypotheses, 500~\GeV, 1000~\GeV\ and 2000~\GeV, 
  in the resolved analysis. The values are extracted by
  calculating the difference in quadrature between the total
  statistical error and the error obtained after setting constant the normalisation
  factor of the background that dominates the region of interest.}
   \label{tab:pre-fit-stat}
\begin{center}
\begin{tabular}{l|c|c|c|c}
Statistical source & \multicolumn{4}{c}{Resolved analysis} \\
\hline
            & \emph{Non-Res} (\%) &  500~\GeV\ (\%)  & 1000~\GeV\ (\%) & 2000~\GeV\ (\%)\\
\hline
Signal region     & +60/--40 & +60/--60 & +70/--60& +80/--70    \\
Top control region       & +40/--30 & +28/--30 & +20/--12 & +13/--13 \\
Multijet control region & +40/--30 & +24/--26 & +30/--30 & +30/--30\\
\hline
Total statistical & +80/--60 & +70/--70     & +80/--70 & +90/--80  \\

\end{tabular}
\end{center}
\end{table}

\begin{table}
\caption{Systematic contributions  (in percentage) to the total error in
  the scaling factor $\alpha_{\mathrm{sig}}$ for
  the non-resonant signal and three scalar-signal mass hypotheses, 500~\GeV, 1000~\GeV\ and 2000~\GeV, 
  in the resolved analysis. The first column quotes the source
of the systematic uncertainty. The $"-"$ symbol indicates that the specified source
is negligible. The contribution is obtained by calculating the
difference in quadrature between the total error in $\alpha_{\mathrm{sig}}$
and that obtained by setting constant the nuisance parameter(s)  relative to the
contribution(s) under study. }
\label{tab:pre-fit-systematics}
\begin{center}
\begin{tabular}{l*{4}{|r@{}c@{}l}}
Systematic source & \multicolumn{12}{c}{Resolved analysis} \\ 
\hline
  & \multicolumn{3}{c|}{\emph{Non-Res} (\%)} &  \multicolumn{3}{c|}{500~\GeV\ (\%)}  & \multicolumn{3}{c|}{1000~\GeV\ (\%)} & \multicolumn{3}{c}{2000~\GeV\ (\%)} \\
\hline
$t\bar{t}$ modelling ISR/FSR & ~~~~+30&/&--20 &~~~~ +10&/&--5  &~~~~~  +7&/&--4  & ~~~~~+2&/&--2    \\
Multijet uncertainty         & +10&/&--10 & +20&/&--10  & +20&/&--20 & +30&/&--30 \\
$t\bar{t}$ Matrix Element                & +10&/&--10 & \multicolumn{3}{c|}{---}     & \multicolumn{3}{c|}{---}     & \multicolumn{3}{c}{---}     \\
$W+$jets modelling PDF       & +4&/&--7   & +10&/&--10 & +2&/&--6   & +7&/&--5  \\
$W+$jets modelling scale     & +9&/&--10  &  +9&/&--4  & +9&/&--2   & +20&/&--10 \\
$W+$jets modelling gen.    & +10&/&--8  & +10&/&--10 & +9&/&--1   & +9&/&--9  \\
$t\bar{t}$ modelling PS       & +3&/&--2   & +30&/&--20 & +20&/&--20 & +2&/&--2    \\
\hline
$b$-tagging    & +30&/&--20 & +11&/&--5  & +7&/&--6   & +30&/&--30 \\

JES/JER             & +13&/&--20 & +20&/&--20 & +50&/&--50  & +10&/&--6  \\

$E_\text{T}^\text{miss}$ soft term res.& +20&/&--20 & +8&/&--1   & +9&/&--7 &
                                                                     +7&/&--7 \\
Pile-up reweighting           & +3&/&--10 & +5&/&--3   & +9&/&--10  & +6&/&--6 \\
\hline
Total systematic & +60&/&--80 & +70&/&--70     & +&60/&--70 & +40&/&--60  \\
\end{tabular}
\end{center}
\end{table}

\clearpage

\subsection{Boosted analysis}
\label{sec:boosted_results}
The boosted analysis applies the selection criteria described in
Section~\ref{sec:boosted_selection}. After applying the large-$R$ jet mass requirement 
$90 < m_\text{Large-$R$ jet} < 140$~\GeV, the $m_{HH}$ distribution is
reconstructed and its shape is fit to data using MC
signal and background templates. The distribution is fit using 17
bins, with almost uniform width except at low and high $m_{HH}$, where
the bin width is modified in order to have a MC statistical uncertainty
smaller than 20\%. All backgrounds, except multijet, are
simulated using MC generators and normalised using the cross section of the
simulated process. The multijet background is estimated using the ABCD
method, and its normalisation obtained from this method is kept fixed in the fit. The bias due
to possible signal contamination in the ABCD regions was studied
and found to have negligible effect on the result.  The integral of the $m_{HH}$
distribution for the boosted analysis is shown in
Table~\ref{tab:event_yields_high}.
\begin{table}
\caption{Data event yields, and post-fit signal and background event yields in the final signal region for the boosted analysis and the scalar $S$ and graviton ($c=1.0$ and $c=2.0$) $G^{\ast}_{\rm KK}$ particle hypotheses. The errors shown
 are the MC statistical and systematic uncertainties described in
 Section~\ref{sec:boosted_systematics}. For illustration a signal mass point of
 2000~\GeV\ is reported in the table. The signal samples are normalised to the expected upper limit cross sections.}  
\label{tab:event_yields_high}
\small
\begin{center}
\begin{tabular}{c|c|c|c|c|c}
$m_X$ [\GeV] & $S$ & $G^{\ast}_{\rm KK}$ ($c=1.0$) & $G^{\ast}_{\rm KK}$ ($c=2.0$) & Total Bkg. & Data
\vspace{0.2mm}\\
\hline
2000 & 28 $\pm$ 0.5  & 36.4 $\pm$ 0.8 & 43.0 $\pm$ 0.7   & 1255 $\pm$ 27  & 1107 \\
\end{tabular}
\end{center}
\end{table}
 
Systematic uncertainties affecting the $m_{HH}$ shape are parameterised as linear functions of $m_{HH}$,
and the function parameters are treated as nuisance parameters in the fit. Statistical uncertainties due to the limited
sample sizes of the simulated background processes are taken into
account in the fit by means of further nuisance parameters, which are
parameterised by Poisson priors. 

The systematic uncertainties included in the fit are described in
Section~\ref{sec:boosted_systematics}. The contribution of the systematic uncertainties to the
total uncertainty is decomposed into the dominant components and summarised  in
Table~\ref{tab:boosed_nplist_systematics}. The most relevant
systematic uncertainties are due to the limited size of the MC
samples, the $t \bar{t}$ modelling and the $b$-tagging systematic uncertainties.

\begin{table}
\caption{Statistical and systematic contributions (in percentage) to
  the total error in the scaling factor $\alpha_{\mathrm{sig}}$ in the boosted analysis
  for four mass hypotheses: 1500~\GeV, 2000~\GeV, 2500~\GeV\ and 3000~\GeV. 
  The first column quotes the source of the uncertainty. 
  The contribution is obtained by calculating the
difference in quadrature between the total error in $\alpha_{\mathrm{sig}}$
and that obtained by setting constant the nuisance parameter(s)  relative to the
contribution(s) under study.
}
   \label{tab:boosed_nplist_systematics}
\begin{center}
\begin{tabular}{l*{4}{|r@{}c@{}l}}
Uncertainty source & \multicolumn{12}{c}{Boosted analysis} \\ 
\hline
       & \multicolumn{3}{c|}{1500~\GeV\ [\%]} &  \multicolumn{3}{c|}{2000~\GeV\ [\%]}  & \multicolumn{3}{c|}{2500~\GeV\ [\%]} & \multicolumn{3}{c}{3000~\GeV\ [\%]} \\
\hline
Data statistics  & ~~+50&/&--52 & ~~+59&/&--61  &  ~~+64&/&--66  & ~~+70&/&--72    \\
Total systematic  & +87&/&--85 & +81&/&--79  &  +76&/&--75  & +71&/&--69    \\
\hline
\hline
MC statistics  & +42&/&--48 & +42&/&--50  &  +39&/&--48  & +39&/&--49    \\
$t\bar{t}$ modelling  & +29&/&--31 & +36&/&--38  &  +40&/&--45  & +32&/&--39    \\
Multijet uncertainty         & +11&/&--14 & +19&/&--23  & +16&/&--20 & +11&/&--16 \\
$W+$jets modelling       & +27&/&--30   & +8&/&--12 & +11&/&--10   & +11&/&--10  \\
Single-top modelling       & +22&/&--26   & +5&/&--6 & +4&/&--5   & +5&/&--5  \\
\hline
$b$-tagging     & +31&/&--19 & +36&/&--22  & +36&/&--17  & +34&/&--14 \\
JES/JER            & +14&/&--14 & +6&/&--6 & +14&/&--11  & +7&/&--9  \\
Large-$R$ jet         & +29&/&--10 & +27&/&--8 & +27&/&--7  & +29&/&--8  \\

\end{tabular}
\end{center}
\end{table}

\begin{figure}
\begin{center}
\includegraphics[height=0.55\textwidth]{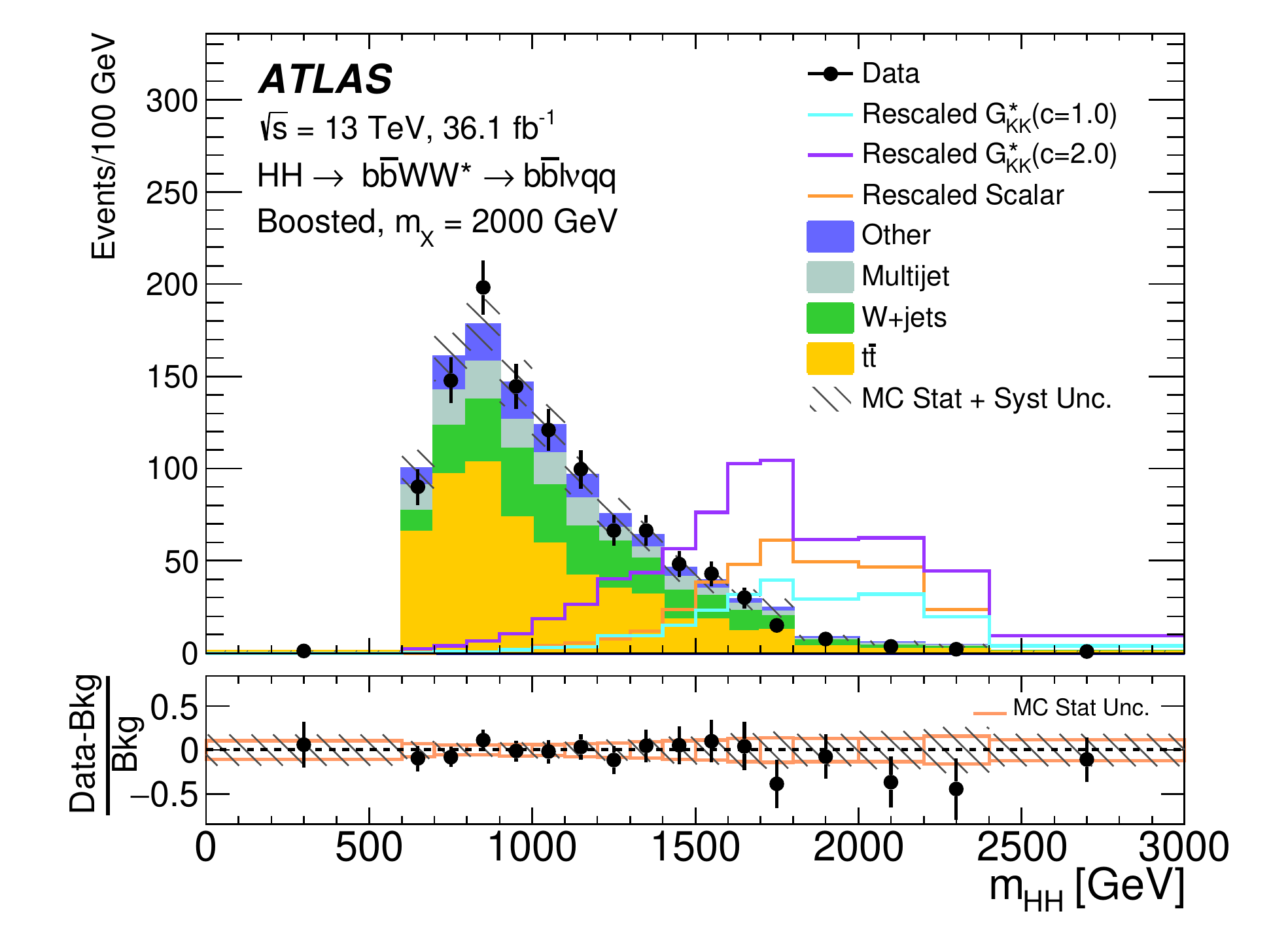} 

\end{center}
\caption{$m_{HH}$ distributions after the global likelihood fit for the boosted analysis. The lower panel shows the fractional difference
   between data and the total expected background with the
   corresponding statistical and total uncertainty.  The signals shown correspond to resonances of mass 2000~\gev. 
   The scalar signal is multiplied by a factor of 4, and both graviton
   signal samples by a factor of 20 with respect to the expected upper-limit 
   cross section reported in Section~\ref{sec:MainResults}.}
   \label{fig:boosted_postfit_mhh}
\end{figure}

\begin{figure}[!h]
\begin{center}
\includegraphics*[width=0.45\textwidth]{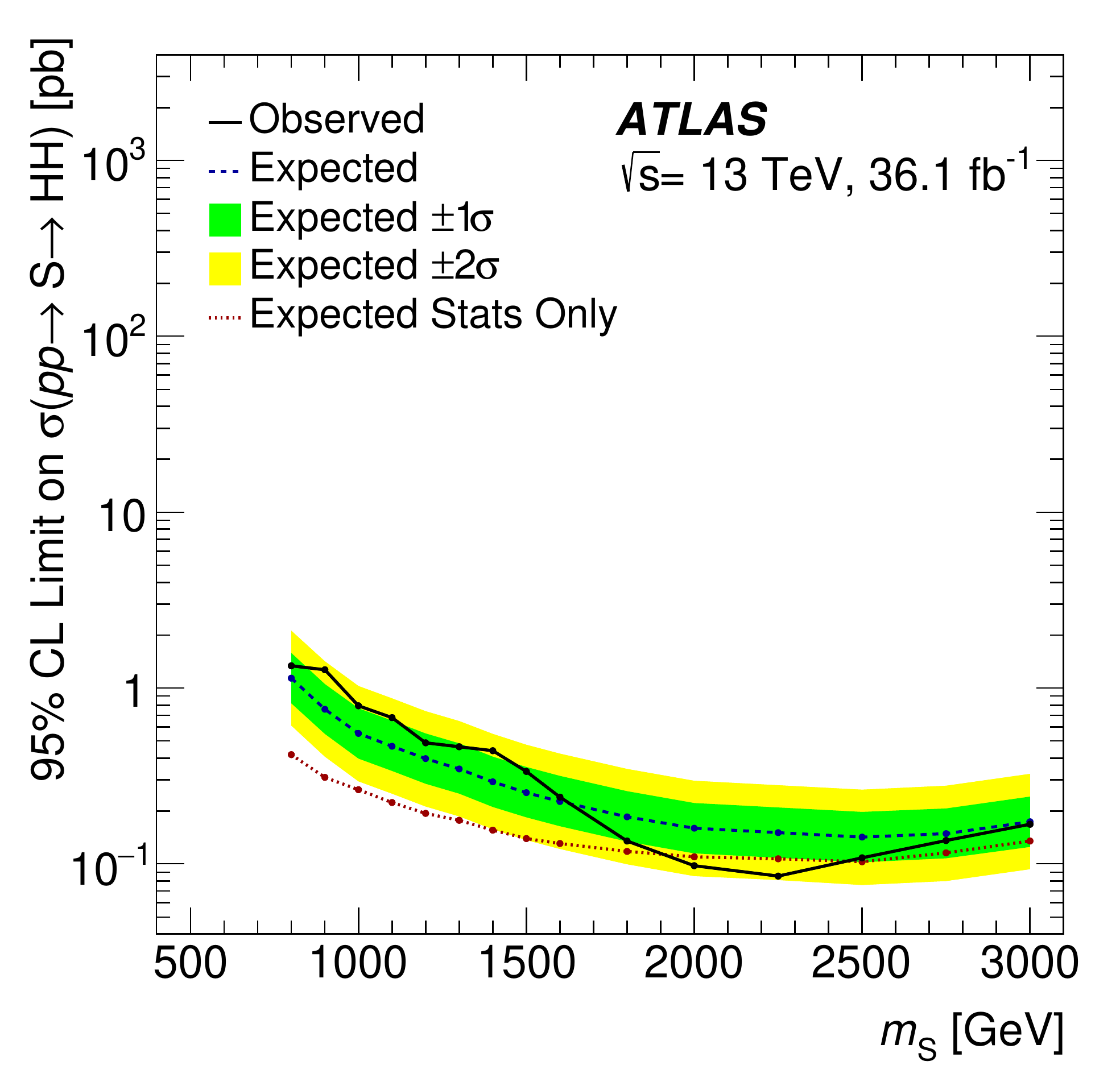}
\includegraphics*[width=0.45\textwidth]{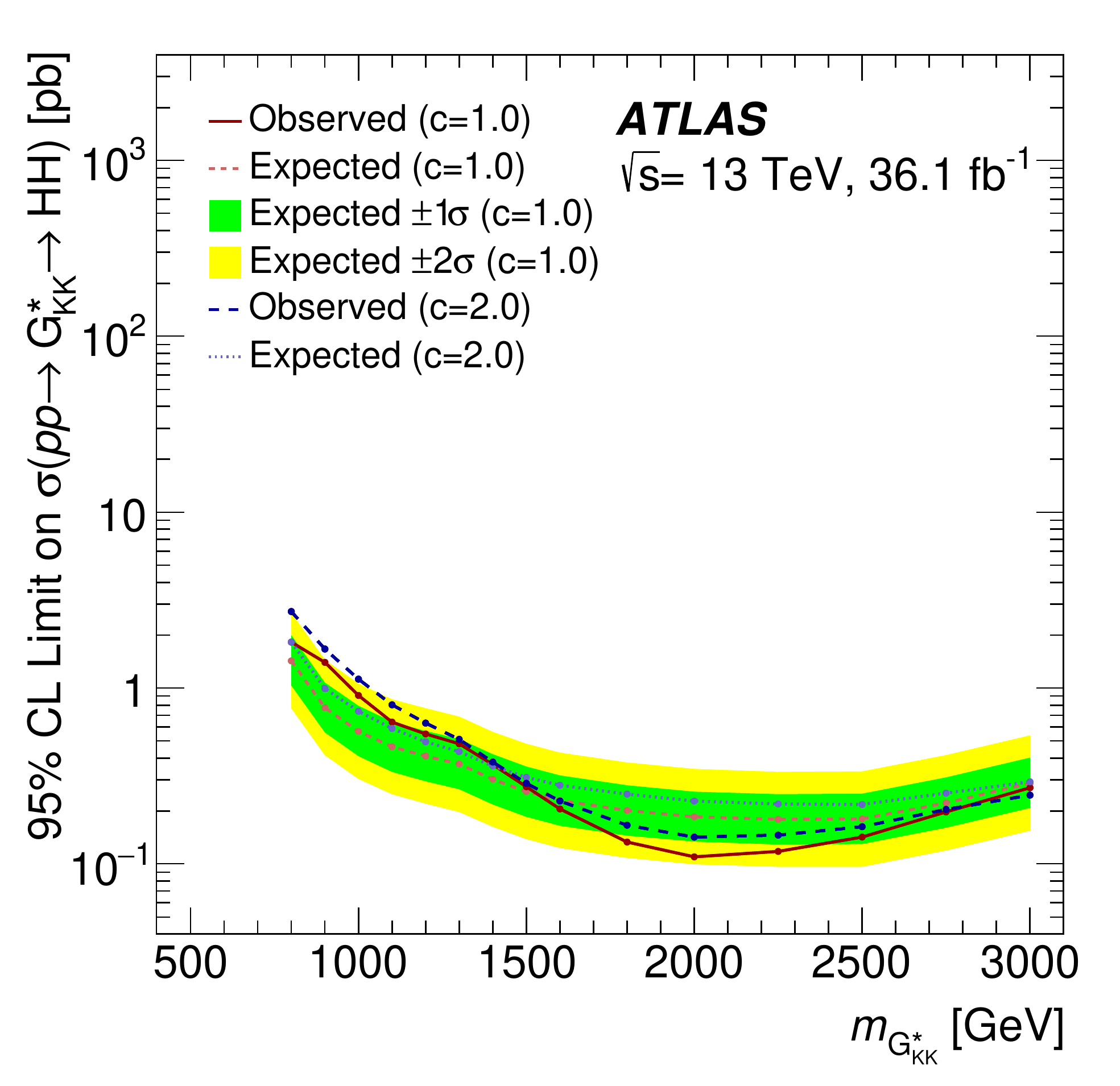}
\caption{Expected and observed upper limits at 95\% CL on the cross section of resonant pair
         production for the heavy scalar boson $S$ model (left) and for
         spin-2 graviton model in two $c$ parameter hypotheses (right) in the boosted analysis. The left plot also shows the expected limits without including the systematic errors in order to show their impact. 
The impact of systematic errors is similar for the graviton models. 
          }

\label{fig:boosted_only_limits}
\end{center}
\end{figure}

Figure~\ref{fig:boosted_postfit_mhh} shows the $m_{HH}$ distribution for
data and the background components for the boosted analysis. 
Data are generally in good agreement with the background expectations within the quoted systematic errors.
The signal $m_{HH}$ distribution is shown in the figure for  the scalar resonance,
and the two graviton hypotheses with $c = 1.0$ and $c=2.0$. 
Figure~\ref{fig:boosted_only_limits} shows the observed and the expected
upper limit on the production cross section of the scalar $S$ and the
two graviton ($c=1.0$ and $c=2.0$)  $G^{\ast}_{\rm KK}$ particles.

\FloatBarrier

\subsection{Summary}
\label{sec:summary_results}

The boosted and resolved analyses have non-trivial event overlap, so their limits are jointly represented through a priority scheme where the two analyses are presented in the mass range where they are most
sensitive.  The boosted analysis was studied for masses larger than
800~\GeV. 
For the graviton interpretation, the boosted analysis is more sensitive than the resolved analysis in the entire studied mass range. For the scalar interpretation, the boosted analysis is more sensitive for $m_{X} > 1300$~\GeV.

\begin{figure}[!h]
\begin{center}
\includegraphics[height=0.41\textwidth]{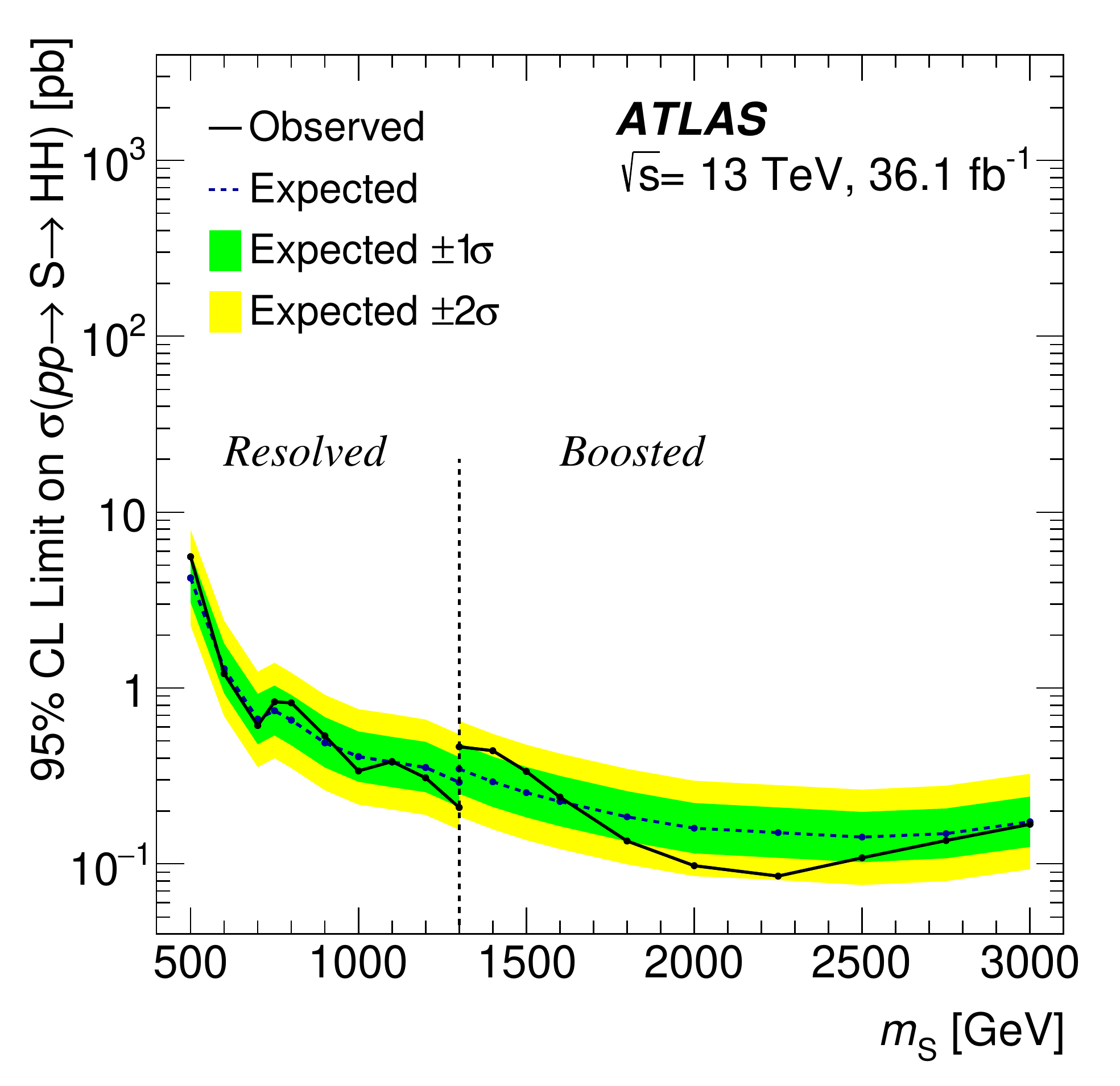} 
\includegraphics*[width=0.4225\textwidth]{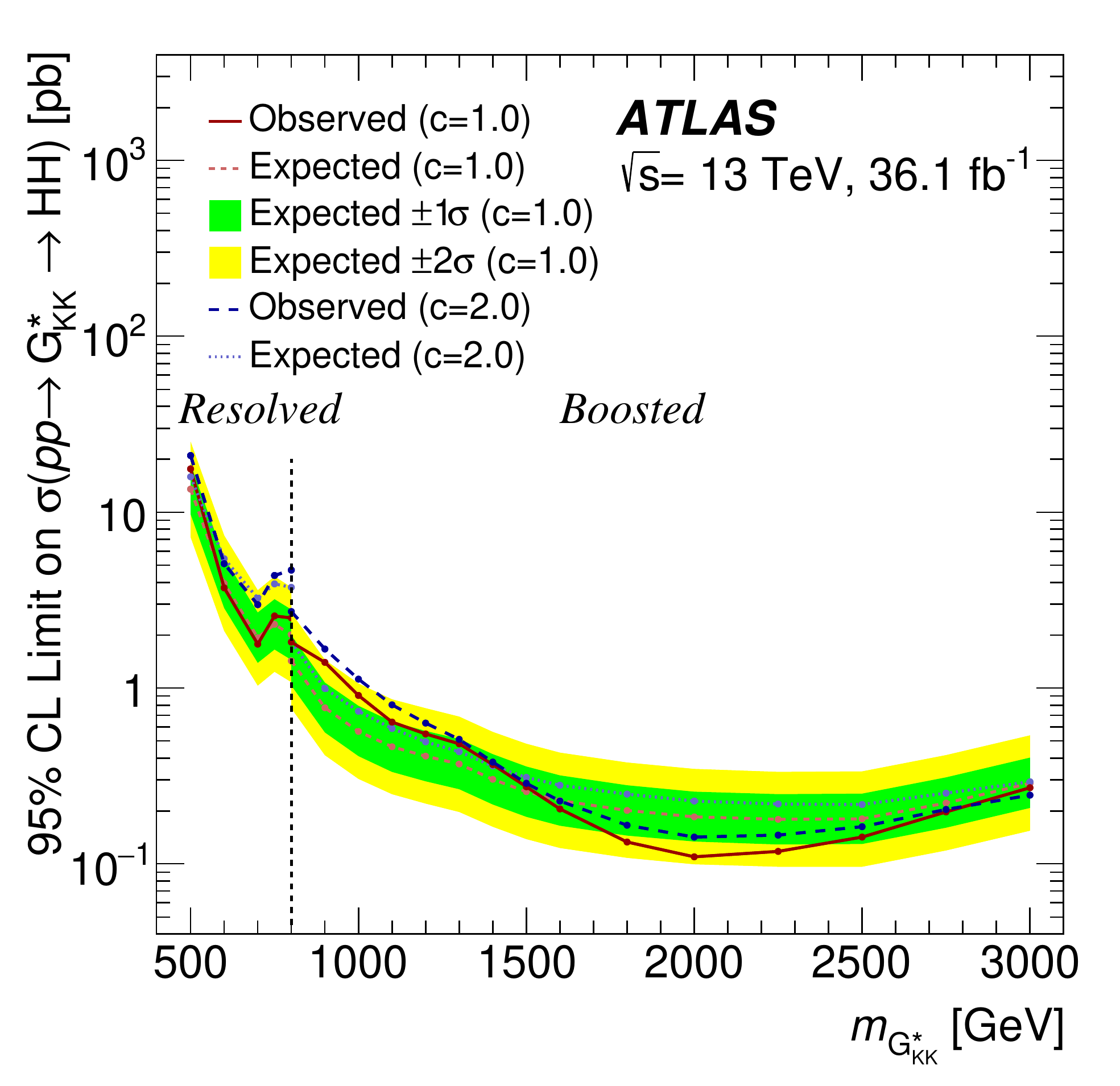}
\end{center}
\caption{Combined plot of 95\% CL cross-section limits for the
  resonant scalar signal model (left) and the
  resonant graviton signal model for the $c=1.0$ and $c=2.0$ hypotheses (right).}
\label{fig:combined_scalar_limit}
\end{figure}

The final result is shown in Figure~\ref{fig:combined_scalar_limit}. The observed upper limits on the production cross sections range from 5.6~pb  for $m_X=500$~\GeV\ to 0.51~pb for 
$m_X=3000$~\GeV\ in the case of a scalar hypothesis. For graviton signals in the same mass interval, they range from 
21~pb (18~pb) to 0.28~pb (0.31~pb) in the case of a hypothesis with $c=2.0$ $(c=1.0)$.  No boosted analysis was
performed for the non-resonant SM signal model.

For the non-resonant signal hypothesis the observed (expected) upper
limit on the $\sigma(pp \to HH) \times {\mathcal{B}}(HH \to bbWW^{\ast})$ at 95\%
CL is:
\[
\sigma(pp \to HH) \cdot {\mathcal{B}}(HH \to bbWW^{\ast}) < 2.5 \, \left
  (2.5^{+1.0}_{-0.7} \right )  \,
{\mathrm{pb}},
\]

which corresponds to 300~(300$^{+100}_{-80}$) times the SM predicted cross section.
\FloatBarrier

\section{Conclusion}
A search for  resonant and non-resonant Higgs boson pair production in
the $b\bar{b}WW^\ast$ decay mode is performed in the $b\bar{b}\ell\nu qq$
final state using $pp$ collision data corresponding to an integrated
luminosity of 36.1~fb$^{-1}$, collected at $\sqrt{s}=13$~\TeV\ by the
ATLAS detector at the Large Hadron Collider. No evidence of an excess of events
over the background expectation is found. 
Limits are set on resonant production as a function of the resonance mass for a scalar resonance and for 
spin-2 gravitons in the mass range 500 to 3000~\GeV.
An upper limit is set on the cross section of non-resonant pair production
$\sigma(pp \to HH) \cdot {\mathcal{B}}(HH \to bbWW^{\ast}) < 2.5$~pb 
at 95\% CL corresponding to  300 times the predicted SM cross
section. Given the result of this work, in order to bring
relevant sensitivity improvement to the $HH$ non-resonant SM searches
in this channel at the LHC and at future colliders, more advanced analysis techniques,
development of new methods for the normalisation of the $t\bar{t}$
background, and a more refined estimation of the multijet background, need to be deployed.

\section*{Acknowledgements}

We thank CERN for the very successful operation of the LHC, as well as the
support staff from our institutions without whom ATLAS could not be
operated efficiently.

We acknowledge the support of ANPCyT, Argentina; YerPhI, Armenia; ARC, Australia; BMWFW and FWF, Austria; ANAS, Azerbaijan; SSTC, Belarus; CNPq and FAPESP, Brazil; NSERC, NRC and CFI, Canada; CERN; CONICYT, Chile; CAS, MOST and NSFC, China; COLCIENCIAS, Colombia; MSMT CR, MPO CR and VSC CR, Czech Republic; DNRF and DNSRC, Denmark; IN2P3-CNRS, CEA-DRF/IRFU, France; SRNSFG, Georgia; BMBF, HGF, and MPG, Germany; GSRT, Greece; RGC, Hong Kong SAR, China; ISF and Benoziyo Center, Israel; INFN, Italy; MEXT and JSPS, Japan; CNRST, Morocco; NWO, Netherlands; RCN, Norway; MNiSW and NCN, Poland; FCT, Portugal; MNE/IFA, Romania; MES of Russia and NRC KI, Russian Federation; JINR; MESTD, Serbia; MSSR, Slovakia; ARRS and MIZ\v{S}, Slovenia; DST/NRF, South Africa; MINECO, Spain; SRC and Wallenberg Foundation, Sweden; SERI, SNSF and Cantons of Bern and Geneva, Switzerland; MOST, Taiwan; TAEK, Turkey; STFC, United Kingdom; DOE and NSF, United States of America. In addition, individual groups and members have received support from BCKDF, CANARIE, CRC and Compute Canada, Canada; COST, ERC, ERDF, Horizon 2020, and Marie Sk{\l}odowska-Curie Actions, European Union; Investissements d' Avenir Labex and Idex, ANR, France; DFG and AvH Foundation, Germany; Herakleitos, Thales and Aristeia programmes co-financed by EU-ESF and the Greek NSRF, Greece; BSF-NSF and GIF, Israel; CERCA Programme Generalitat de Catalunya, Spain; The Royal Society and Leverhulme Trust, United Kingdom. 

The crucial computing support from all WLCG partners is acknowledged gratefully, in particular from CERN, the ATLAS Tier-1 facilities at TRIUMF (Canada), NDGF (Denmark, Norway, Sweden), CC-IN2P3 (France), KIT/GridKA (Germany), INFN-CNAF (Italy), NL-T1 (Netherlands), PIC (Spain), ASGC (Taiwan), RAL (UK) and BNL (USA), the Tier-2 facilities worldwide and large non-WLCG resource providers. Major contributors of computing resources are listed in Ref.~\cite{ATL-GEN-PUB-2016-002}.

\clearpage
\printbibliography
\clearpage
 
\begin{flushleft}
{\Large The ATLAS Collaboration}

\bigskip

M.~Aaboud$^\textrm{\scriptsize 34d}$,    
G.~Aad$^\textrm{\scriptsize 99}$,    
B.~Abbott$^\textrm{\scriptsize 124}$,    
O.~Abdinov$^\textrm{\scriptsize 13,*}$,    
B.~Abeloos$^\textrm{\scriptsize 128}$,    
D.K.~Abhayasinghe$^\textrm{\scriptsize 91}$,    
S.H.~Abidi$^\textrm{\scriptsize 164}$,    
O.S.~AbouZeid$^\textrm{\scriptsize 39}$,    
N.L.~Abraham$^\textrm{\scriptsize 153}$,    
H.~Abramowicz$^\textrm{\scriptsize 158}$
% ...
\bigskip
\\

% ...
$^{13}$Institute of Physics, Azerbaijan Academy of Sciences, Baku; Azerbaijan.\\
% ...
$^{34}$$^{(a)}$Facult\'e des Sciences Ain Chock, R\'eseau Universitaire de Physique des Hautes Energies - Universit\'e Hassan II, Casablanca;$^{(b)}$Centre National de l'Energie des Sciences Techniques Nucleaires (CNESTEN), Rabat;$^{(c)}$Facult\'e des Sciences Semlalia, Universit\'e Cadi Ayyad, LPHEA-Marrakech;$^{(d)}$Facult\'e des Sciences, Universit\'e Mohamed Premier and LPTPM, Oujda;$^{(e)}$Facult\'e des sciences, Universit\'e Mohammed V, Rabat; Morocco.\\
% ...
$^{39}$Niels Bohr Institute, University of Copenhagen, Copenhagen; Denmark.\\
% ...
$^{91}$Department of Physics, Royal Holloway University of London, Egham; United Kingdom.\\
% ...
$^{99}$CPPM, Aix-Marseille Universit\'e, CNRS/IN2P3, Marseille; France.\\
% ...
$^{124}$Homer L. Dodge Department of Physics and Astronomy, University of Oklahoma, Norman OK; United States of America.\\
% ...
$^{128}$LAL, Universit\'e Paris-Sud, CNRS/IN2P3, Universit\'e Paris-Saclay, Orsay; France.\\
% ...
$^{153}$Department of Physics and Astronomy, University of Sussex, Brighton; United Kingdom.\\
% ...
$^{158}$Raymond and Beverly Sackler School of Physics and Astronomy, Tel Aviv University, Tel Aviv; Israel.\\
% ...
$^{164}$Department of Physics, University of Toronto, Toronto ON; Canada.\\
% ...
$^{a}$ Also at Borough of Manhattan Community College, City University of New York, NY; United States of America.\\
$^{b}$ Also at California State University, East Bay; United States of America.\\
$^{c}$ Also at Centre for High Performance Computing, CSIR Campus, Rosebank, Cape Town; South Africa.\\
$^{d}$ Also at CERN, Geneva; Switzerland.\\
$^{e}$ Also at CPPM, Aix-Marseille Universit\'e, CNRS/IN2P3, Marseille; France.\\
% ...
$^{*}$ Deceased

\end{flushleft}


\end{document}